\documentclass[twocolumn,pra,footinbib,superscriptaddress]{revtex4-2}

\usepackage[utf8]{inputenc}

\usepackage{amsmath}
\usepackage{amsthm}
\usepackage{amssymb}
\usepackage{graphicx}

\usepackage{color}
\usepackage{braket}
\usepackage{epstopdf}
\usepackage{dcolumn}
\usepackage{textcomp}

\usepackage[
colorlinks=true,        % 开启彩色链接（false则为带边框的无色链接）
linkcolor=red,         % 内部链接（如章节、公式引用）颜色
anchorcolor=black,      % 锚点颜色
citecolor=red,        % 参考文献引用颜色
urlcolor=blue,       % 外部网址链接颜色
filecolor=cyan,         % 文件链接颜色
pdfstartview=FitH,      % PDF打开时的视图（FitH=适配宽度）
pdfpagemode=UseNone,    % PDF打开时不显示书签/缩略图
bookmarks=true,         % 生成PDF书签（侧边栏）
bookmarksnumbered=true, % 书签显示章节编号
unicode=true,           % 支持中文书签/链接
breaklinks=true         % 自动换行长链接
]{hyperref}

\usepackage{url}

\makeatother

\begin{document}
\title{Giant atoms coupled to waveguide: Continuous coupling and multiple
excitations}
\author{Shiying Lin}
\author{Xinyu Zhao}
\email{xzhao@fzu.edu.cn}

\author{Yan Xia}
\email{xia-208@163.com}

\affiliation{Fujian Key Laboratory of Quantum Information and Quantum Optics (Fuzhou
University), Fuzhou 350116, China}
\affiliation{Department of Physics, Fuzhou University, Fuzhou 350116, China}
\begin{abstract}
We propose a stochastic Schrödinger equation (SSE)
approach to investigate the dynamics of giant atoms coupled to a waveguide,
addressing two critical gaps in existing research, namely insufficient
exploration on continuous coupling and multiple excitations. A key
finding is that continuous coupling, unlike discrete coupling at finite
points, breaks the constant phase difference condition, thereby weakening
the interference effects in giant atom-waveguide systems. In addition,
a key technical advantage of the SSE approach is that auto- and cross-correlation
functions can directly reflect the complex photon emission/absorption
processes and time-delay effects in giant atom-waveguide systems.
Moreover, the SSE approach also naturally handles multiple excitations,
without increasing equation complexity as the number of excitations
grows. This feature enables the investigation of multi-excitation
initial states of the waveguide, such as thermal and squeezed initial
states. Overall, our approach provides a powerful tool for studying
the dynamics of giant atoms coupled to waveguide, particularly for
continuous coupling and multi-excitation systems.
\end{abstract}
\maketitle

\section{Introduction}\label{sec:Introduction}

In recent years, artificial giant atoms have attracted the attention
of many researchers due to their unique properties~\cite{Kannan2020,PhysRevA.106.013715,PhysRevResearch.2.043070,PhysRevLett.130.053601,PhysRevA.95.053821,Anderson2019,PhysRevLett.122.203603,PhysRevA.105.023712,Leonforte2025,PhysRevA.106.063717,PhysRevA.90.013837,PhysRevA.101.053855,PhysRevLett.128.223602,Anderson2019,Du2023,Cheng2022,Weng2024}.
Since the size of giant atoms cannot be ignored, the coupling between
giant atoms and light fields can be more complicated~\cite{PhysRevA.109.023720,PhysRevA.105.023712,PhysRevA.104.013720}.
In particular, when giant atoms are coupled to a waveguide, a series
of novel physical phenomena emerge due to interference and time-delay
effects that do not occur in small atom systems \cite{PhysRevLett.120.140404,Yu2025OE,Qiu2025NJP,Xu2025OE,Xiao2024PRA,Xiao2023AQT,Yan2024PRA,Yan2025OE,Shi2024PRAa,Shi2024PRA,Chen2024PRL,Zheng2023PRA,Li2024PRA}.
For example, the complicated interaction between waveguide and giant
atoms leads to collective enhancement and interference \cite{qiu_2023_collective,PhysRevA.109.023712,PhysRevA.109.033711},
frequency-dependent Lamb shift and relaxation~\cite{PhysRevA.90.013837},
single-photon scattering~\cite{PhysRevA.106.013715,PhysRevA.101.053855,PhysRevA.104.063712,PhysRevA.104.023712,chen_2022_nonreciprocal,PhysRevA.107.063703},
and decoherence-free interaction~\cite{PhysRevLett.120.140404,PhysRevResearch.2.043184,PhysRevA.107.023705,PhysRevA.105.023712,PhysRevA.107.013710}.~Additionally,
it also induces non-Markovian decay dynamics~\cite{PhysRevA.107.023705,PhysRevA.102.033706,PhysRevA.108.013704},
novel bound states~\cite{PhysRevResearch.2.043014,PhysRevA.101.053855,PhysRevLett.126.043602,PhysRevA.104.053522,PhysRevA.107.023716},
phase regulation and photon shielding~\cite{PhysRevA.101.053855,PhysRevLett.121.153601}.

\begin{figure}[htbp]
\centering \includegraphics[width=1\columnwidth]{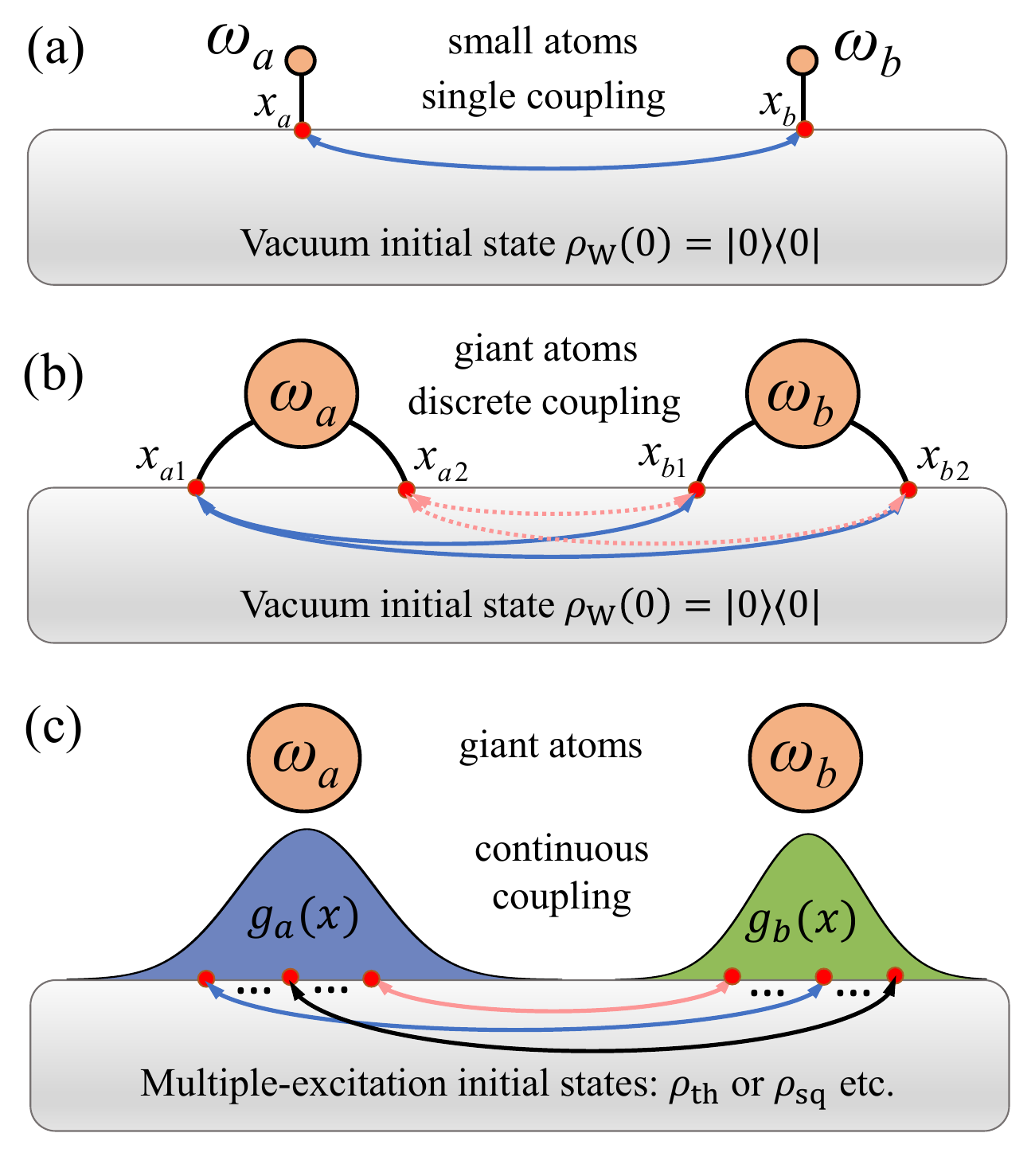} \caption{Schematic diagram of the two giant atoms (labeled
\textquotedblleft$a$\textquotedblright{} and \textquotedblleft$b$\textquotedblright )
coupled to a waveguide. (a) Single coupling (small atom) with vacuum
initial state of the waveguide. (b) Discrete coupling (two coupling
points) with vacuum initial state of the waveguide. (c) Continuous
coupling with multiple-excitation (e.g. thermal or squeezed) initial
states of the waveguide.}
\label{fig:1} 
\end{figure}

All these novel phenomena related to giant atoms
fundamentally originate from the interference effects in the photon
emission and absorption process. \cite{PhysRevA.104.033710,PhysRevA.106.013715,PhysRevA.109.023712,PhysRevA.109.063703,PhysRevA.109.033711}.
As shown in Fig.~\ref{fig:1}(a), in the case of small atoms coupled
to waveguide, photons emitted by an atom can be only absorbed by other
atoms at fixed positions, with deterministic phase accumulation \cite{PhysRevA.88.043806,PhysRevA.100.013812}.
In contrast, as shown in Fig.~\ref{fig:1}(c), giant atoms enable
photon emission from arbitrary point in a continuous region, followed
by potential reabsorption at arbitrary point in a continuous region~\cite{PhysRevLett.120.140404,PhysRevA.103.023710,PhysRevA.108.023728,PhysRevA.106.063703,PhysRevA.104.033710,PhysRevA.104.063712,Li:23,PhysRevA.106.063717}.
This complex photon emission and absorption process
greatly enriches the interference effects among various possible paths.~\cite{Cheng2023,PhysRevA.102.033706,PhysRevResearch.2.043014,Kannan2020,PhysRevA.95.053821}.
However, two critical gaps remain in existing research, limiting our
understanding of giant atom-waveguide dynamics.

Gap 1: Rarely explored continuous coupling

Existing studies have predominantly focused on two (or a few) discrete
coupling points as shown in Fig.~\ref{fig:1}(b)~\cite{PhysRevA.101.053855,PhysRevLett.133.063603,PhysRevLett.120.140404,Zhu2025},
leaving the novel physics induced by continuous coupling
{[}Fig.~\ref{fig:1}(c){]} rarely discussed.

As we have discussed, all interesting phenomena in
giant atom systems stem from interference effects \cite{Du2023,Kannan2020,Leonforte2025,PhysRevA.109.033711,PhysRevA.109.023720}.
When there are only a few discrete coupling {[}(e.g., two coupling
points as shown in Fig.~\ref{fig:1}(b){]}, the phase difference
is essentially fixed, leading to strong constructive and destructive
interference. In the case of continuous coupling points shown in Fig.~\ref{fig:1}(c),
however, this condition of constant phase difference is broken, which
inevitably weakens the interference effects. Exploring how the broadening
of the coupling region influences the dynamics of giant atoms constitutes
a valuable research topic.

Gap 2: Rarely addressed multiple excitation effects

In many studies, the dynamical equations are obtained
by assuming single excitation in the initial state to limit the Hilbert
space to finite dimensions~\cite{PhysRevA.106.063703,PhysRevLett.130.053601,PhysRevLett.128.223602,PhysRevLett.133.063603}
(see discussion in Sec.~\ref{sec:SingleExcitation}). But the multiple
excitation effect \cite{Yin2025} is rarely discussed, particularly
for the multi-photon initial states in the waveguide.

Notably, waveguides support a complete continuum
of modes, including ultra-low-frequency modes with vanishingly small
energies. At finite temperatures, the thermal occupation of these
low-frequency modes becomes non-negligible. Therefore, the single-excitation
assumption is invalid in practical cases. Hence, studying multiple
excitation effects is essential to capture the realistic dynamics
of giant atoms.

In this paper, inspired by the non-Markovian quantum state diffusion
method in the theory of open quantum systems \cite{PhysRevA.58.1699,PhysRevLett.82.1801,PhysRevA.60.91,chen_2022_calculating,gao_2019_charge,PhysRevA.85.042106,PhysRevA.84.032101,PhysRevA.86.032116,ZHAO2017121,Zhao:19,Zhao2022,Zhao2025OE,Xiang2025PRA,Xiang2025LP},
we propose an alternative approach, the stochastic Schrödinger equation
(SSE) approach, to derive the dynamical equations governing the time
evolution of giant atoms. By expanding the quantum states of the waveguide
in coherent state basis, we derive the SSE and the corresponding master
equation from the original Hamiltonian without introducing phenomenological
treatments. The proposed theoretical framework enables the investigation
beyond single-excitation state, while simultaneously offering complete
characterization of multiple-excited initial states including thermal~\cite{PhysRevA.69.062107}
and squeezed states~\cite{PhysRevA.108.012206} in the waveguide.

With the SSE approach, we systematically investigate
how the continuous coupling affects the quantum dynamics of giant
atoms in a waveguide. A notable advantage of the SSE approach is that
the correlation functions intuitively capture photon emission and
absorption processes, which serves as a powerful tool to analyze the
time-delayed effects. Our numerical results show that continuous coupling
with a single-peak Gaussian distribution enables more photon emission
and absorption paths, thereby enhancing of entanglement generation.

As the most important result of this work, it should
be highlighted that continuous coupling between giant atoms and the
waveguide over a region (rather than discrete coupling at several
specific points) leads to complex photon emission-absorption processes
with diverse propagation times, which breaks the constant phase difference
condition. Since this condition is essential for interference effects,
continuous coupling may compromise the performance of most established
applications of giant atoms relying on interference.

Finally, we extend our analysis beyond the single-excitation
subspace. For double-excitation states, we analyze the influence of
the distribution width on entanglement dynamics and explain the reason
through photon emission and absorption picture. Last but not least,
we demonstrate that our approach is applicable to the multi-photon
initial states, including thermal and squeezed initial states of the
waveguide.

\section{The model: Giant atoms coupled to waveguide}

\label{sec:II}

As shown in Fig.\,\ref{fig:1}(c), we consider two giant atoms coupled
to a waveguide as an example. The total Hamiltonian (setting $\hbar=1$)
is given by 
\begin{equation}
H_{\text{tot}}=H_{\text{A}}+H_{\text{W}}+H_{\text{int}},\label{eq:Htot}
\end{equation}
\begin{equation}
H_{\text{A}}=\sum_{\mu=a,b}\omega_{\mu}\sigma_{\mu}^{+}\sigma_{\mu}^{-},\label{eq:HA}
\end{equation}
\begin{equation}
\begin{aligned}H_{\text{W}}=\sum_{k}\omega_{k}c_{k}^{\dagger}c_{k},\end{aligned}
\label{eq:HW}
\end{equation}
\begin{equation}
H_{\text{int}}=\sum_{\mu=a,b}\sum_{k}\left(G_{\mu k}c_{k}^{\dagger}\sigma_{\mu}^{-}+{\rm H.c.}\right),\label{eq:Hint}
\end{equation}
where $H_{\text{A}}$, $H_{\text{W}}$, and $H_{\text{int}}$ indicate
the Hamiltonian for two giant atoms, the waveguide, and their interaction,
respectively. The transition frequencies for two giant atoms ``$a$''
and ``$b$'' are described by $\omega_{\mu}$ ($\mu=a,b$), and
$\omega_{k}$ is the frequency of the $k^{{\rm th}}$ mode in the
waveguide. In most of the numerical simulation we often take $\omega_{a}=\omega_{b}=\omega$.
The operators $\sigma_{\mu}^{+}$ and $\sigma_{\mu}^{-}$ represent
the raising and lowering operators, and $c_{k}$ is the photon annihilation
operator.

Equation (\ref{eq:Hint}) describes the interaction between two giant
atoms and a waveguide in frequency domain. From the perspective of
spatial distribution, the coupling between atoms and the waveguide
can be characterized by a spatial distribution function $g_{\mu}(x)$.
The frequency-dependent coupling strength $G_{\mu k}$~is given by
a Fourier transformation \cite{PhysRevResearch.6.013279} 
\begin{equation}
G_{\mu k}=\int dxg_{\mu}(x)e^{-ikx}.\label{eq:Guk}
\end{equation}
In the limiting case that the size of the atoms is extremely small,
$g_{\mu}(x)$ is reduced to $\delta$-function as shown in Fig.~\ref{fig:1}(a),
the model is reduced to the small atom case, since each atom only
couples to the waveguide at a single point. In this case, the photon
emission and absorption is the simplest. Photon emitted from one atom
must be absorbed by the other one at the fixed position.

In most existing studies about giant atoms \cite{PhysRevA.106.013715,PhysRevA.109.023720,PhysRevA.104.013720,qiu_2023_collective,PhysRevA.109.023712,PhysRevLett.120.140404},
researchers mainly focus on the case that giant atoms are coupled
to the waveguide with two or finite coupling points as shown in Fig.~\ref{fig:1}(b).
In this case, photon emitted, for instance, at $x_{a1}$ can be absorbed
at different positions. But the choice is still limited. It is worth
to note that the case of two coupling points is sufficient to capture
most of the interesting physical phenomena arising from photon emission
and absorption, although the photon emission and absorption are still
simple. In this paper, we consider the most general case as shown
in Fig.~\ref{fig:1}(c), the distribution function $g_{\mu}(x)$
is a continuous distribution in the space domain. This enables more
complicated photon emission and absorption process.

\section{Limitation of the conventional single-excitation approach}\label{sec:SingleExcitation}

The most widely used method for solving the dynamics of giant atom
systems is the dressed state method, in which the Hilbert space is
often limited in single (or a few) excitation subspace \cite{PhysRevA.106.063703,PhysRevA.106.013702,PhysRevLett.130.053601,PhysRevLett.128.223602}.
For example, in the model described in Eq.~(\ref{eq:Htot}), the
total excitation number operator $N=\sum_{\mu=a,b}\sigma_{\mu}^{+}\sigma_{\mu}^{-}+\sum_{k}c_{k}^{\dagger}c_{k}$
is a conserved quantity, then an arbitrary quantum state in the single-excitation
subspace of the system can be expressed as 
\begin{equation}
|\psi^{(1)}(t)\rangle=\sum_{\mu=a,b}A_{\mu}(t)e^{-i\omega_{\mu}t}\sigma_{\mu}^{+}|G\rangle+\sum_{k}B_{k}(t)e^{-i\omega_{k}t}c_{k}^{\dagger}|G\rangle,\label{eq:SingleExcitationState}
\end{equation}
with $\sum_{\mu=a,b}|A_{\mu}(t)|^{2}+\sum_{k}|B_{k}(t)|^{2}=1$, where
$|G\rangle=|g_{a}\rangle|g_{b}\rangle|0\rangle$ represents the waveguide
is in the vacuum state and the giant atoms are in their ground states.
The $A_{\mu}(t)$ is the probability amplitude of the atom ``$\mu$'',
and $B_{k}(t)$ denotes the single-photon probability amplitude of
the mode $k$. Based on the Schrödinger equation $\frac{d}{dt}|\psi^{(1)}(t)\rangle=-iH_{{\rm tot}}|\psi^{(1)}(t)\rangle$,
one can obtain the equations of motion for these probability amplitudes~\cite{PhysRevA.106.063703}
\begin{equation}
\begin{aligned}\dot{A}_{\mu}(t) & =-i\sum_{k}G_{\mu k}^{*}B_{k}(t)e^{-i(\omega_{k}-\omega_{\mu})t},\\
\dot{B}_{k}(t) & =-i\sum_{\mu=a,b}G_{\mu k}A_{\mu}(t)e^{i(\omega_{k}-\omega_{\mu})t}.
\end{aligned}
\label{eq:dressstate}
\end{equation}
Subsequently, the system dynamics can be found by solving above differential
equations.

However, the limitation of this method is obvious. When the excitation
number is high, both the number and complexity of the equations will
increase significantly (see Sec.~\ref{subsec:6.1}). Besides,
the waveguide supports a continuum of modes, including those with
vanishingly low frequencies ($\omega_{k}\rightarrow0$).~Even near
absolute zero, these low-frequency modes maintain a non-negligible
thermal population $\bar{n}_{k}=1/(e^{\beta\hbar\omega_{k}}-1)$.
Consequently, at any finite temperature ($T>0$), the waveguide is
inherently in a multi-excitation state. This invalidates the single-excitation
assumption. Thus, the limitation of this conventional
approach demands a new framework.

\section{Stochastic Schr\"{o}dinger equation approach}\label{sec:SSEA}

\subsection{Equation for the stochastic state vector}\label{subsec:SSE}

To overcome the limitations of the single-excitation
approximation, we propose the SSE approach based on the non-Markovian
quantum state diffusion approach~\cite{PhysRevA.58.1699,PhysRevLett.82.1801,PhysRevA.60.91,chen_2022_calculating,gao_2019_charge,PhysRevA.85.042106,PhysRevA.84.032101,PhysRevA.86.032116,ZHAO2017121,Zhao:19,Zhao2022,Zhao2025OE}.
Next, we start from the original Hamiltonian (1) to derive the SSE
for the two giant atoms, as well as the master equation.

To focus on the dynamics of the giant atoms, the
reduced density operator is defined by performing a partial trace
over the waveguide degrees of freedom 
\begin{equation}
\rho(t)={\rm tr_{W}}\left[|\psi_{\text{tot}}(t)\rangle\langle\psi_{\text{tot}}(t)|\right].\label{eq:trace}
\end{equation}
Evaluating this trace in the Fock basis is the most intuitive approach,
yielding 
\begin{equation}
\rho(t)=\sum_{\{n_{k}\}}\langle\{n_{k}\}|\psi_{\text{tot}}(t)\rangle\langle\psi_{\text{tot}}(t)|\{n_{k}\}\rangle,\label{traceFock}
\end{equation}
where $|\{n_{k}\}\rangle\equiv\prod\otimes_{k}|n_{k}\rangle$ denotes
the multi-mode Fock state of the waveguide, with each mode $k$ in
the Fock state $|n_{k}\rangle$. However, since the waveguide contains
infinite modes, resulting in a prohibitively large Hilbert space,
making this Fock-basis trace evaluation computationally intractable.

To overcome this difficulty, we transform the partial
trace into a stochastic averaging problem by using the Bargmann coherent
states $|z\rangle\equiv\prod\otimes_{k}|z_{k}\rangle$, with each
mode $k$ in the coherent state $|z_{k}\rangle$. The properties of
coherent states (e.g., $c_{k}|z\rangle=z_{k}|z\rangle$) facilitate
the subsequent derivation of the dynamical equation. Substituting
the completeness relation $\int\frac{d^{2}z}{\pi}e^{-|z|^{2}}|z\rangle\langle z|=1$
into the trace operation in Eq.~(\ref{traceFock}), the reduced density
operator can be rewritten as 
\begin{align}
\rho(t) & =\int\frac{d^{2}z}{\pi}e^{-|z|^{2}}\langle z|\psi_{\text{tot}}(t)\rangle\langle\psi_{\text{tot}}(t)|z\rangle\nonumber \\
 & \equiv\mathcal{M}[|\psi(t,z^{*})\rangle\langle\psi(t,z)|],\label{eq:Mean}
\end{align}
where the stochastic state vector is defined as 
\begin{equation}
|\psi(t,z^{*})\rangle\equiv\langle z|\psi_{\text{tot}}(t)\rangle,\label{eq:psi}
\end{equation}
and the stochastic average is defined as $\mathcal{M}[\cdot]\equiv\int\frac{d^{2}z}{\pi}e^{-|z|^{2}}[\cdot]$
\cite{PhysRevA.58.1699,PhysRevA.60.91}. Equation (\ref{eq:Mean})
clarifies the physical meaning of the stochastic state vector $|\psi(t,z^{*})\rangle$.
For a particular set of parameters $\{z_{k}\}$, $|\psi(t,z^{*})\rangle$
represents a possible trajectory. The statistical average over all
possible trajectories will reproduce the density operator $\rho(t)$
\cite{PhysRevA.58.1699,PhysRevA.60.91}.

Next, we will derive the equation governing the evolution
of the stochastic state vector $|\psi(t,z^{*})\rangle$. As we know,
the total state vector $|\psi_{\text{tot}}(t)\rangle$ satisfies the
Schrödinger equation $\frac{d}{dt}|\psi_{\text{tot}}(t)\rangle=-iH_{\text{tot}}^{I}(t)|\psi_{\text{tot}}(t)\rangle$,
where 
\begin{equation}
H_{\text{tot}}^{I}(t)=H_{\text{A}}+\sum_{\mu=a,b}\left(\sum_{k}G_{\mu k}\sigma_{\mu}^{-}c_{k}^{\dagger}e^{i\omega_{k}t}+{\text{H}.{\rm c}.}\right),
\end{equation}
is the Hamiltonian in the interaction picture with respect to $H_{\text{W}}$.~Taking
time derivative to both sides of Eq.~(\ref{eq:psi}) and substituting
the Schrödinger equation, one obtains 
\begin{equation}
\begin{aligned}\frac{d}{dt}|\psi(t,z^{*})\rangle=\bigg[ & -iH_{\text{A}}+\sum_{\mu=a,b}\Big(\sigma_{\mu}^{-}z_{\mu t}^{*}\\
 & -i\sigma_{\mu}^{+}\sum_{k}G_{\mu k}^{*}e^{-i\omega_{k}t}\frac{\partial}{\partial z_{k}^{*}}\Big)\bigg]|\psi(t,z^{*})\rangle,
\end{aligned}
\label{eq:Schrodinger}
\end{equation}
where $z_{\mu t}^{*}\equiv-i\sum_{k}G_{\mu k}z_{k}^{*}e^{i\omega_{k}t}$
is a $z_{k}$-dependent function. Given a particular set of $z_{k}^{*}$,
there will be a realization of $z_{\mu t}^{*}$ and a corresponding
solution $|\psi(t,z^{*})\rangle$ of Eq.~(\ref{eq:Schrodinger}).
Diósi~et~al.~\cite{PhysRevA.58.1699} named $z_{\mu t}^{*}$ as
a stochastic noise and the corresponding solution $|\psi(t,z^{*})\rangle$
is called a quantum trajectory. In principle, the density operator
can be retrieved by integrating over all possible trajectories as
shown in Eq.~(\ref{eq:Mean}). However, in practical scenarios, there
is no need to generate an infinite number of trajectories. Only a
few hundred (at most a few thousand) randomly generated trajectories
are enough to reproduce density operator with high precision. This
approach replaces the cumbersome trace operation in Eq.~(\ref{eq:trace})
with a stochastic average over trajectories, thereby greatly simplifying
the computation.

Equation~(\ref{eq:Schrodinger}) is merely a formal
dynamical equation. It contains both $z_{\mu t}^{*}$ and $\frac{\partial}{\partial z_{k}^{*}}$.
To obtain a computationally tractable equation which only depends
on $z_{\mu t}^{*}$ and the time $t$, one can use the chain rule
to replace the term $\frac{\partial}{\partial z_{k}^{*}}|\psi(t,z^{*})\rangle$
by a functional derivative $-i\sum_{k}G_{\mu k}\sigma_{\mu}^{+}e^{-i\omega_{k}t}\frac{\partial}{\partial z_{k}^{*}}=-\sigma_{\mu}^{\dagger}\int_{0}^{t}ds[\sum_{k}\frac{\partial z_{\mu t}}{\partial z_{k}}(\frac{\partial z_{\mu s}^{*}}{\partial z_{k}^{*}}\frac{\delta}{\delta z_{\mu s}^{*}}+\frac{\partial z_{\nu s}^{*}}{\partial z_{k}^{*}}\frac{\delta}{\delta z_{\nu s}^{*}})]$
$(\mu,\nu=a,b;\mu\neq\nu)$. Then, Eq.~(\ref{eq:Schrodinger}) becomes
\begin{widetext}
\begin{equation}
\frac{d}{dt}|\psi(t,z^{*})\rangle=\sum_{\substack{\mu,\nu=a,b\\
\mu\ne\nu
}
}\left\{ -iH_{\text{A}}+\sigma_{\mu}^{-}z_{\mu t}^{*}-\sigma_{\mu}^{+}\int_{0}^{t}ds\left[\alpha_{\mu\mu}(t,s)\frac{\delta}{\delta z_{\mu s}^{*}}+\alpha_{\mu\nu}(t,s)\frac{\delta}{\delta z_{\nu s}^{*}}\right]\right\} |\psi(t,z^{*})\rangle.\label{eq:S}
\end{equation}
where 
\begin{equation}
\alpha_{\mu\mu}(t,s)=\sum_{k}\frac{\partial z_{\mu t}}{\partial z_{k}}\frac{\partial z_{\mu s}^{*}}{\partial z_{k}^{*}}=\sum_{k}{|G_{\mu k}|^{2}e^{-i\omega_{k}(t-s)}},\label{eq:CFuu}
\end{equation}
are the auto-correlation functions, and 
\begin{equation}
\alpha_{\mu\nu}(t,s)=\sum_{k}\frac{\partial z_{\mu t}}{\partial z_{k}}\frac{\partial z_{\nu s}^{*}}{\partial z_{k}^{*}}=\sum_{k}{G_{\mu k}^{*}G_{\nu k}e^{-i\omega_{k}(t-s)}},\label{eq:CFuv}
\end{equation}
are the cross-correlation functions for subscripts $\mu\neq\nu$.
Equation~(\ref{eq:S}) is directly derived from Schrödinger equation
and includes stochastic noises $z_{\mu t}^{*}$, thus it is called
the SSE \cite{PhysRevA.58.1699,PhysRevA.60.91,PhysRevLett.82.1801}.
\end{widetext}

It is important to note that the correlation functions
$\alpha_{\mu\mu}$ and $\alpha_{\mu\nu}$ encode all information about
the interactions between giant atoms and waveguide. We emphasize that
these functions are predetermined by the physical setup described
by the spatial coupling distribution $g_{\mu}(x)$. They are independent
of the state of the giant atoms. Actually, the correlation functions
can also be defined in terms of the interaction Hamiltonian, which
is presented in Appendix~\ref{sec:AppCF}. This confirms from another
perspective that the coupling between the waveguide and the giant
atoms can be encoded in the correlation functions. It is straightforward
to check the correlation functions characterize the statistics of
noise $z_{\mu t}^{*}$, as evidenced by the following relations 
\begin{equation}
\begin{aligned}\mathcal{M}[z_{\mu t}]=\mathcal{M} & [z_{\mu t}z_{\mu s}]=0,\\
\mathcal{M}[z_{\mu t}z_{\mu s}^{*}]=\alpha_{\mu\mu}(t,s),\  & \mathcal{M}[z_{\mu t}z_{\nu s}^{*}]=\alpha_{\mu\nu}(t,s).
\end{aligned}
\end{equation}
In the numerical simulation, one need to generate stochastic functions
$z_{\mu t}^{*}$ that satisfies these correlation functions \cite{Zhao2022}.
The detailed discussion about the correlation functions is presented
in Sec.~\ref{sec:Continuous}.

Up to this point, we have derived SSE in Eq.~(\ref{eq:S}).
Starting from nothing more than the Hamiltonian, we expand the state
vector $|\psi_{{\rm tot}}\rangle$ into many stochastic state vectors
$|\psi(t,z^{*})\rangle$ using the coherent state basis and derived
corresponding equation governing the dynamics of $|\psi(t,z^{*})\rangle$
(SSE). Solving the SSE under different noise realizations $z_{\mu t}^{*}$
produces numerous distinct trajectories $|\psi(t,z^{*})\rangle$.
Averaging these trajectories as $\mathcal{M}[|\psi(t,z^{*})\rangle\langle\psi(t,z)|]$
yields the density operator $\rho(t)$. Throughout the derivation,
we only employ the Schrödinger equation, thus the SSE is purely a
mathematical result.

Besides, our Hamiltonian contains two different coupling
$G_{\mu k}$, thus we introduce two noises $z_{\mu t}^{\ast}\,(\mu=a,b)$,
resulting in two types of correlation functions, auto- and cross-correlation
functions. This is an extension of the original SSE approach~\cite{PhysRevA.58.1699,PhysRevLett.82.1801,PhysRevA.60.91}.
In our study of giant atoms, a by-product emerged, that the framework
of the SSE approach is extended to the case involving two types of
noises and correlation functions.

\subsection{Mathematical Techniques for solving SSE}

The primary challenge in solving Eq.~(\ref{eq:S}) stems from the
functional derivatives $\frac{\delta}{\delta z_{\mu s}^{*}}|\psi(t,z^{*})\rangle$.
To address this difficulty, we introduce time-dependent operators
$O_{\mu}$ ($\mu=a,b$) defined as 
\begin{equation}
\frac{\delta}{\delta z_{\mu s}^{*}}|\psi(t,z^{*})\rangle=O_{\mu}(t,s,z^{*})|\psi(t,z^{*})\rangle.
\end{equation}
With the operators $O_{\mu}$, Eq.~(\ref{eq:S}) is rewritten as
\begin{equation}
\begin{gathered}\frac{d}{dt}|\psi(t,z^{*})\rangle=H_{{\rm eff}}|\psi(t,z^{*})\rangle\\
H_{{\rm eff}}=-iH_{\text{A}}+\sum_{\mu=a,b}\left[\sigma_{\mu}^{-}z_{\mu t}^{*}-\sigma_{\mu}^{+}\bar{O}_{\mu}(t,z^{*})\right]
\end{gathered}
\label{eq:NMQSD}
\end{equation}
where the operators $\bar{O}_{\mu}$ are defined as 
\begin{equation}
\bar{O}_{\mu}(t,z^{*})=\int_{0}^{t}ds\sum_{\nu=a,b}\left[\alpha_{\mu\nu}(t,s)O_{\nu}(t,s,z^{*})\right].
\end{equation}
For simplicity, we use the abbreviations $O_{\mu}\equiv O_{\mu}(t,s,z^{*})$
and $\bar{O}_{\mu}\equiv\bar{O}_{\mu}(t,z^{*})$ if no confusion arises.

Now, if the solution of $O_{\mu}$ can be found, Eq.~(\ref{eq:S})
can be solved as an ordinary differential equation with stochastic
variables. Actually, the operators $O_{\mu}$ can be determined through
the consistency condition \cite{PhysRevA.60.91}
\begin{equation}
\frac{\partial}{\partial t}\frac{\delta}{\delta z_{\mu s}^{*}}|\psi(t,z^{*})\rangle=\frac{\delta}{\delta z_{\mu s}^{*}}\frac{\partial}{\partial t}|\psi(t,z^{*})\rangle.\label{eq:CC}
\end{equation}
Substituting Eq.~(\ref{eq:NMQSD}) into Eq.~(\ref{eq:CC}), the
left-hand-side (LHS) yields
\begin{widetext}
\begin{equation}
\begin{aligned}\frac{\partial}{\partial t}\frac{\delta}{\delta z_{\mu s}^{*}}|\psi(t,z^{*})\rangle & =\left(\frac{\partial}{\partial t}O_{\mu}\right)|\psi(t,z^{*})\rangle+O_{\mu}\frac{\partial}{\partial t}|\psi(t,z^{*})\rangle=\left(\frac{\partial}{\partial t}O_{\mu}+O_{\mu}H_{{\rm eff}}\right)|\psi(t,z^{*})\rangle,\end{aligned}
\label{eq:LHS}
\end{equation}
and the right-hand-side (RHS) yields
\begin{equation}
\begin{aligned}\frac{\delta}{\delta z_{\mu s}^{*}}\frac{\partial}{\partial t}|\psi(t,z^{*})\rangle & =\left(\frac{\delta}{\delta z_{\mu s}^{*}}H_{{\rm eff}}\right)|\psi(t,z^{*})\rangle+H_{{\rm eff}}\left(\frac{\delta}{\delta z_{\mu s}^{*}}|\psi(t,z^{*})\rangle\right)=\left(H_{{\rm eff}}O_{\mu}-\sum_{\nu=a,b}\sigma_{\nu}^{+}\frac{\delta}{\delta z_{\mu s}^{*}}\bar{O}_{\nu}\right)|\psi(t,z^{*})\rangle.\end{aligned}
\label{eq:RHS}
\end{equation}
Equating LHS and RHS, one can obtain

\begin{align}
\frac{\partial}{\partial t}O_{\mu} & =\sum_{\nu=a,b}\Bigg\{\left[-iH_{\text{A}}+\sigma_{\nu}^{-}z_{\nu t}^{*}-\sigma_{\nu}^{+}\bar{O}_{\nu},O_{\mu}\right]-\sigma_{\nu}^{+}\frac{\delta}{\delta z_{\mu s}^{*}}\bar{O}_{\nu}\Bigg\},\label{eq:dtO}
\end{align}
\end{widetext}

subject to the initial conditions $O_{\mu}(t,s=t,z^{*})=\sigma_{\mu}^{-}$.
In this model, it is straightforward to obtain the solution for $O_{\mu}$
is \cite{PhysRevA.84.032101} 
\begin{equation}
O_{\mu}(t,s)=\sum_{j=1}^{4}p_{\mu j}(t,s)O_{\mu j},\label{eq:Oab}
\end{equation}
where the basis operators are 
\begin{align}
 & O_{\mu1}=\sigma_{\mu}^{-},\ O_{\mu2}=\sigma_{\mu}^{z}\sigma_{\nu}^{-},\nonumber \\
 & O_{\mu3}=\sigma_{\nu}^{-},\ O_{\mu4}=\sigma_{\nu}^{z}\sigma_{\mu}^{-}.\label{eq:O}
\end{align}
According to Eq.~(\ref{eq:dtO}), the time-dependent coefficients
$p_{\mu j}$ are governed by the coupled partial differential equations
\begin{widetext}
\begin{equation}
\begin{aligned}\frac{\partial p_{\mu1}}{\partial t}= & i\omega_{\mu}p_{\mu1}+p_{\mu1}P_{\mu1}+p_{\mu1}Q_{\nu3}+p_{\mu1}Q_{\mu2}+p_{\mu1}P_{\nu4}-p_{\mu2}Q_{\mu1}+p_{\mu2}Q_{\mu4}\\
 & +p_{\mu2}P_{\nu2}-p_{\mu2}P_{\nu3}+p_{\mu4}P_{\mu4}+p_{\mu4}Q_{\nu2}+p_{\mu4}Q_{\mu2}+p_{\mu4}P_{\nu4},\\
\frac{\partial p_{\mu2}}{\partial t}= & i\omega_{\nu}p_{\mu2}+p_{\mu2}P_{\mu4}+p_{\mu2}Q_{\nu2}+p_{\mu2}Q_{\mu3}+p_{\mu2}P_{\nu1}+p_{\mu1}P_{\mu2}-p_{\mu1}P_{\mu3}\\
 & -p_{\mu1}Q_{\nu1}+p_{\mu1}Q_{\nu4}+p_{\mu3}P_{\mu4}+p_{\mu3}Q_{\nu2}+p_{\mu3}Q_{\mu2}+p_{\mu3}P_{\nu4},\\
\frac{\partial p_{\mu3}}{\partial t}= & i\omega_{\nu}p_{\mu3}+p_{\mu3}P_{\nu1}+p_{\mu3}Q_{\mu3}+p_{\mu3}Q_{\nu2}+p_{\mu3}P_{\mu4}-p_{\mu4}Q_{\nu1}+p_{\mu4}Q_{\nu4}\\
 & +p_{\mu4}P_{\mu2}-p_{\mu4}P_{\mu3}+p_{\mu2}P_{\mu4}+p_{\mu2}Q_{\nu2}+p_{\mu2}Q_{\mu2}+p_{\mu2}P_{\nu4},\\
\frac{\partial p_{\mu4}}{\partial t}= & i\omega_{\mu}p_{\mu4}+p_{\mu4}P_{\mu1}+p_{\mu4}Q_{\nu3}+p_{\mu4}Q_{\mu2}+p_{\mu4}P_{\nu4}+p_{\mu1}P_{\mu4}+p_{\mu1}Q_{\nu2}\\
 & +p_{\mu1}Q_{\mu2}+p_{\mu1}P_{\nu4}-p_{\mu3}Q_{\mu1}+p_{\mu3}Q_{\mu4}+p_{\mu3}P_{\nu2}-p_{\mu3}P_{\nu3},
\end{aligned}
\end{equation}
\end{widetext}

with the initial conditions 
\begin{equation}
p_{\mu1}(t,s=t)=1,\quad p_{\mu j}(t,s=t)=0\ (j\neq1).
\end{equation}
The coefficients $P_{\mu j}$ and $Q_{\mu j}$ are defined as $P_{\mu j}(t)=\int_{0}^{t}ds\alpha_{\mu\mu}(t,s)p_{\mu j}(t,s)$
and $Q_{\mu j}(t)=\int_{0}^{t}ds\alpha_{\nu\mu}(t,s)p_{\mu j}(t,s)$
with $j=1,2,3,4$. It is clear that $P_{\mu j}$ and $Q_{\mu j}$
are time domain convolutions for auto- and cross-correlation functions,
representing the cumulative effect of the impacts exerted by all past
moments $s$ on the present moment $t$.

\subsection{Derivation of master equation from SSE}

With the solution in Eq.~(\ref{eq:Oab}), Eq.~(\ref{eq:NMQSD})
can be solved numerically. However, to further obtain the density
operator $\rho(t)$, one needs to solve Eq.~(\ref{eq:NMQSD}) repeatedly
and obtain a large amount of trajectories of $|\psi(t,z^{*})\rangle$
with different noise realizations $z_{\mu t}^{*}$. The density operator
$\rho$ is then obtained by taking the statistical average over these
trajectories, as shown in Eq.~(\ref{eq:Mean}).

Alternatively, one may also derive the master equation
based on Eq. (\ref{eq:NMQSD}). Following the method established in
\cite{PhysRevA.69.052115,Chen2014PRA}, the master equation can be
derived as (see Appendix~\ref{sec:AppMEQ} for details) 
\begin{equation}
\frac{d}{dt}\rho=-i[H_{\text{A}},\rho]+\sum_{\mu=a,b}\left([\sigma_{\mu}^{-},\rho\bar{O}_{\mu}^{\dagger}]+{\rm H.c.}\right).\label{eq:MEQ}
\end{equation}

Notably, when the correlation functions $\alpha_{\mu\nu}(t,s)$
$(\mu,\nu=a,b)$ is reduced to delta-functions $\alpha_{\mu\nu}(t,s)=\Gamma\delta(t,s)$,
it is straightforward to check all the coefficients $P$ and $Q$
are all zero except $P_{\mu1}=\Gamma/2$ and $Q_{\mu1}=\Gamma/2$.
Thus, $\bar{O}_{\mu}=\Gamma/2(\sigma_{\mu}^{-}+\sigma_{\nu}^{-})$
and the master equation is reduced to the Lindblad master equation
commonly used in the references \cite{PhysRevA.104.013720,PhysRevLett.130.053601}
\begin{equation}
\frac{d}{dt}\rho=-i[H_{\text{A}},\rho]+\sum_{\substack{\mu,\nu=a,b\\
\mu\ne\nu
}
}\frac{\Gamma}{2}\left\{ \left[\sigma_{\mu}^{-},\rho(\sigma_{\mu}^{+}+\sigma_{\nu}^{+})\right]+{\rm H.c.}\right\} .
\end{equation}
This limiting case provides an important connection to conventional
quantum optical treatment.

\section{Entanglement dynamics with continuous coupling}\label{sec:Continuous}

Based on the SSE approach presented in Sec.~\ref{sec:SSEA}, we will
investigate entanglement between two giant atoms from another perspective
and reveal the physical meaning of the correlation functions, which
is rarely touched in existing literature. Throughout our analysis,
we quantify entanglement by using the well-established entanglement
measure, ``concurrence'' $C$\ \cite{PhysRevLett.80.2245}.

\subsection{Transition from discrete coupling to continuous
coupling}\label{subsec:Transition}

\begin{figure*}
\centering \includegraphics[width=1\linewidth]{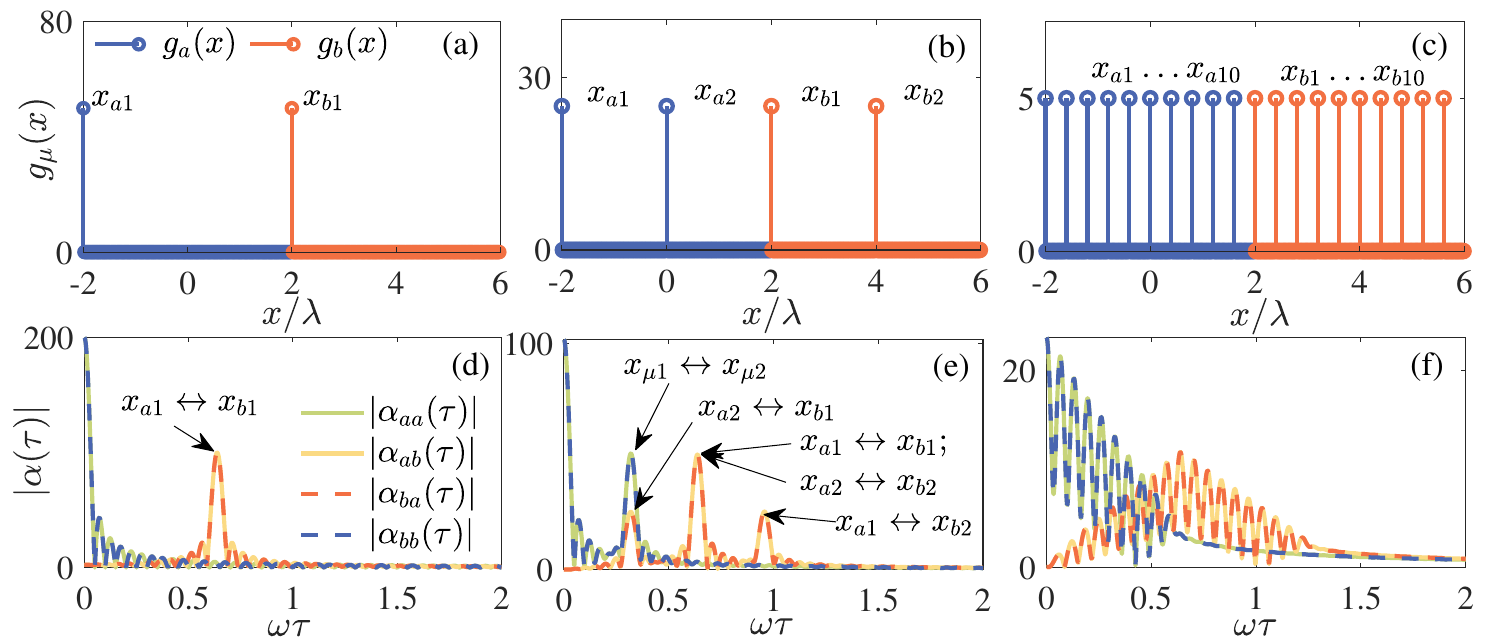} \caption{(a)-(c) Discrete coupling distributions for $m=1$, $m=2$, and $m=10$
coupling points, respectively. (d)-(f) Auto- and cross-correlation
functions for $m=1$, $m=2$, and $m=10$, respectively.}
\label{fig:2} 
\end{figure*}

We start from a comb-like discrete coupling distribution. The comb-like
distribution function is defined as 
\begin{equation}
g_{\mu}(x)=\frac{1}{\tilde{N}}\sum_{i=1}^{m}\delta(x-x_{\mu i}),\quad(\mu=a,b)
\end{equation}
where $\tilde{N}$ denotes the normalization factor, while $x_{\mu i}$
corresponds to the spatial coordinates of non-zero coupling points,
$m$ represents the number of coupling points. Such a distribution
describes the giant atoms are coupled to waveguide only at several
discrete points $x=x_{\mu i}$, illustrated in Fig.~\ref{fig:2}(a)-(c).
The distance is measured in units of the wavelength $\lambda=c/\omega$,
where $c$ is the speed of light (setting to be 1) and $\omega=\omega_{a}=\omega_{b}$
is the transition frequency of two atoms (setting to be identical).
Similarly, time is scaled by the dimensionless quantity $\omega\tau$.
All the figures throughout the paper are presented in this set of
dimensionless units.

Before investigating the time evolution of the quantum
states, we first explore how correlation functions vary with the number
of coupling points. The corresponding correlation functions for each
distribution are displayed in Fig.~\ref{fig:2}(d)--(f). As shown
in Eq.~(\ref{eq:CFuu}) and Eq.~(\ref{eq:CFuv}), the correlation
functions are independent of the initial state and is determined solely
by the distribution of coupling.

For single coupling point ($m=1$) in Fig.~\ref{fig:2}(a)~and~(d),
the cross-correlation function, for example $\alpha_{ab}(\tau)$,
exhibits a peak in the time domain. This indicates that only event
(e.g., photon emitted at $x_{a}$) occurring at a specific moment
$s=t-\tau$ in the past can exert an influence on the present event
(e.g., photon absorbed at $x_{b}$) at time $t$, and this particular
moment precisely corresponds to the time required for a photon to
propagate from point $x_{a}$ to point $x_{b}$.

For two coupling points ($m=2$), the auto-correlation
function $\alpha_{\mu\mu}(\tau)$ in subplot~(e) has one peak (except
the one at $\tau=0$) that corresponds to the physical process that
photon emitted at $x_{\mu1}$ (or $x_{\mu2}$) is absorbed at $x_{\mu2}$
(or $x_{\mu1}$). This allows a time delayed self-absorption at another
coupling point for the same giant atom, which does not exist in the
case of small atom. Besides, the cross-correlation function $\alpha_{\mu\nu}(\tau)$
exhibits three peaks corresponding to three possible emission and
absorption routes: $x_{a2}\leftrightarrow x_{b1}$, $x_{a1}\leftrightarrow x_{b1}$
($x_{a2}\leftrightarrow x_{b2}$), and $x_{a1}\to x_{b2}$, which
have been marked in Fig.~\ref{fig:2}(e).

Extending to more coupling points ($m=10$) shown
in Fig.~\ref{fig:2}(c)~and~(f), there are more feasible pathways
for photon emission and absorption, which collectively constitute
a highly sophisticated emission-absorption network. In Fig.~\ref{fig:2}(f),
both the auto- and cross-correlation functions exhibit multiple peaks,
which reflects that there exist more possibilities in the time duration
for photons to undergo emission and absorption.

As can be seen from Fig.~\ref{fig:2}, within the
framework of the SSE approach, the auto- and cross-correlation functions
can, to some extent, reveal the impact of the delay effect on system
dynamics. Notably, the calculation of the correlation functions depends
solely on the distribution and strengths of coupling points. Consequently,
we can gain preliminary insight into the role of the delay effect
in dynamical evolution by solely considering the distribution pattern
and coupling strengths. This is a feature not possessed by conventional
methods in Sec.~\ref{sec:SingleExcitation}.

\begin{figure}[tb]
\centering \includegraphics[width=1\columnwidth]{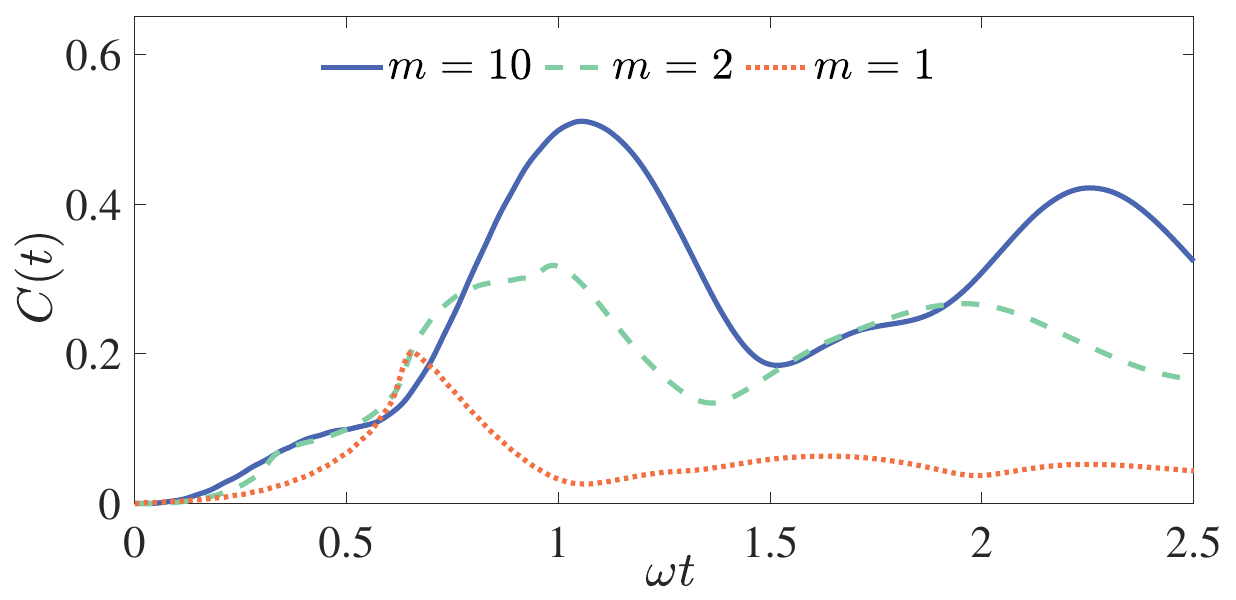} \caption{Time evolution of concurrence $C(t)$ for single-point (orange dotted),
two-point (green dashed), and multi-point (blue solid) coupling cases.
The parameters are identical to those in Fig.~\ref{fig:2}.}
\label{fig:3} 
\end{figure}

\begin{figure*}[t]
\centering \includegraphics[width=1\linewidth]{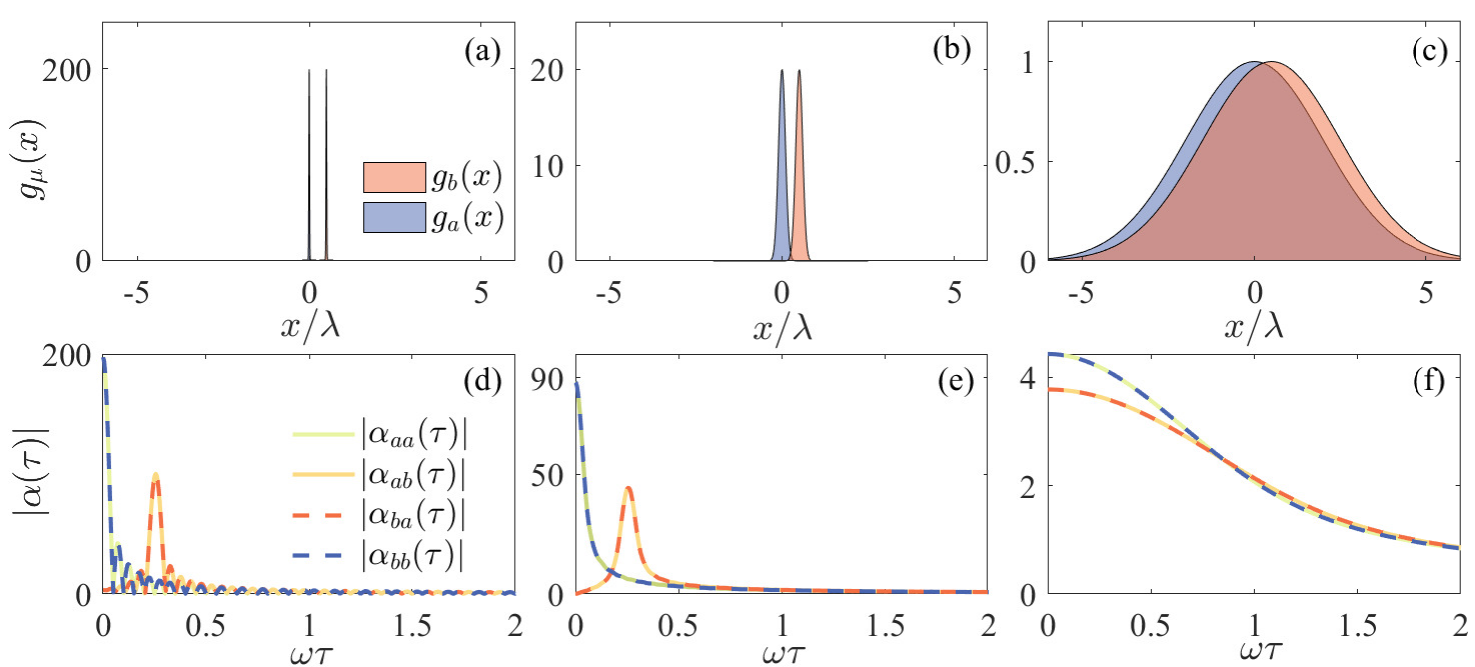} \caption{Gaussian distribution $g_{\mu}(x)$ for $s_{\mu}=0.01$, 0.1, and
2 (assuming $s_{a}=s_{b}$) are plotted in panels (a), (b), (c).~The
correlation functions $\alpha_{\mu\nu}(\tau)$ for $s_{\mu}=0.01$,
$0.1$, and $2$ (assuming $s_{a}=s_{b}$) are plotted in panels (d),
(e), (f).}
\label{fig:4}
\end{figure*}
Having established that the correlation functions
reflect the paths of photon emission and absorption, we next investigate
how the number of coupling points influences the entanglement between
the giant atoms. By choosing the initial state as $|eg\rangle$, the
time evolution of the concurrence is plotted in Fig. \ref{fig:3}.
In the case of a single coupling point ($m=1$, orange dotted line
in Fig.~\ref{fig:3}), the generation of entanglement is transient
and weak ($C_{\max}\approx0.2$), owing to the limited interaction
pathways. When the number of coupling points is increased (e.g., $m=2$
in Fig.~\ref{fig:3}) leads to a greater diversity of pathways for
photon emission and absorption. Such multi-path interference results
in enhanced entanglement generation, yielding a maximum entanglement
of $C_{\max}\approx0.3$. As the number of coupling points continues
to increase ($m=10$ in Fig.~\ref{fig:3}), the peak value of entanglement
$C(t)$ increases significantly to 0.5 with a stronger revival. This
behavior demonstrates that the multiple coherent feedback channels
within the waveguide enable stronger and robust entanglement generation.

\subsection{Continuous coupling with Gaussian distribution}

\label{subsec:Gaussian}

When the coupling points become sufficiently dense, the discrete distribution
transitions into a continuous distribution. We assume the spatially
dependent coupling distribution $g_{\mu}(x)$ is a Gaussian distribution,
\begin{equation}
g_{\mu}(x)=\frac{1}{s_{\mu}\sqrt{2\pi}}\exp\left[-\frac{(x-\bar{x}_{\mu})^{2}}{2s_{\mu}^{2}}\right],\quad(\mu=a,b)\label{eq:GaussianDis}
\end{equation}
where $\bar{x}_{\mu}$ denote the central positions and $s_{\mu}$
characterize the distribution widths. The influence of the distribution
width $s_{\mu}$ is presented in Fig.~\ref{fig:4}. One may compare
the results in Fig.~\ref{fig:4} with those in Fig.~\ref{fig:2}.
In the case of continuous coupling {[}except the extreme localization
case depicted in Fig.~\ref{fig:4}(a){]}, the correlation functions
in Fig.~\ref{fig:4}(e) and Fig.~\ref{fig:4}(f) evolve into continuous
functions instead of exhibiting distinct peaks. This implies that
events occurring at any arbitrary past time $s$ can exert an influence
on the current time $t$, rather than only those events at specific
past times $s$ (satisfying $\ensuremath{t-s=\tau}_{peak}$) as observed
in Fig.~\ref{fig:2}. The fundamental reason is that in the case
of continuous coupling, photons can be emitted from any coupling points
and subsequently absorbed at any other coupling points, such that
the propagation time involved can take on any arbitrary value. In
contrast, under discrete coupling, photons can only be emitted and
absorbed between a limited set of specific coupling points, which
restricts their propagation time to a discrete set of values.

Further discussions on the effects of continuous
coupling on entanglement generation, particularly the impact of the
central positions $\bar{x}_{\mu}$ of the continuous coupling, are
discussed in Appendix~\ref{sec:AppC} and Appendix~\ref{appendix:CenterPosition}.
We now proceed to present the most important results of this work.

\subsection{Breaking constant phase difference via continuous coupling}\label{subsec:break}

As mentioned in Sec.~\ref{sec:Introduction}, all
intriguing phenomena in giant-atom systems stem from interference
effects associated with photon emission and absorption processes.
However, continuous coupling may break the constant phase difference
condition, which consequently weaken such interference effects, as
illustrated in Fig.~\ref{fig:1}(c). To show how continuous coupling
impact interference effects, we consider the spatial distribution
of the continuous coupling $g_{\mu}(x)$ as a double-peak function
\begin{widetext}

\begin{equation}
g_{\mu}(x)=\frac{1}{s_{\mu}\sqrt{2\pi}}\left\{ \exp\left[-\frac{(x-\bar{x}_{\mu1})^{2}}{2s_{\mu}^{2}}\right]+\exp\left[-\frac{(x-\bar{x}_{\mu2})^{2}}{2s_{\mu}^{2}}\right]\right\} .\quad(\mu=a,b)\label{eq:gx_2peak}
\end{equation}

\end{widetext}

This choice is motivated by the connection of this
double-peak spatial distribution to the widely studied discrete coupling
model with two coupling points {[}see Fig.~\ref{fig:1}(b){]}. When
the distribution width $s_{\mu}\rightarrow0$, the double-peak distribution
in Eq.~(\ref{eq:gx_2peak}) reduces to the well-established two-coupling-point
configuration in Fig.~\ref{fig:1}(b). The interference phenomena
have been extensively investigated in previous studies \cite{PhysRevA.107.023705,PhysRevA.106.063703,PhysRevA.108.023728}.
Varying $s_{\mu}$ enables systematic exploration of how continuous
coupling broadening modulates interference effects.

\begin{figure}
\centering \includegraphics[width=1\columnwidth]{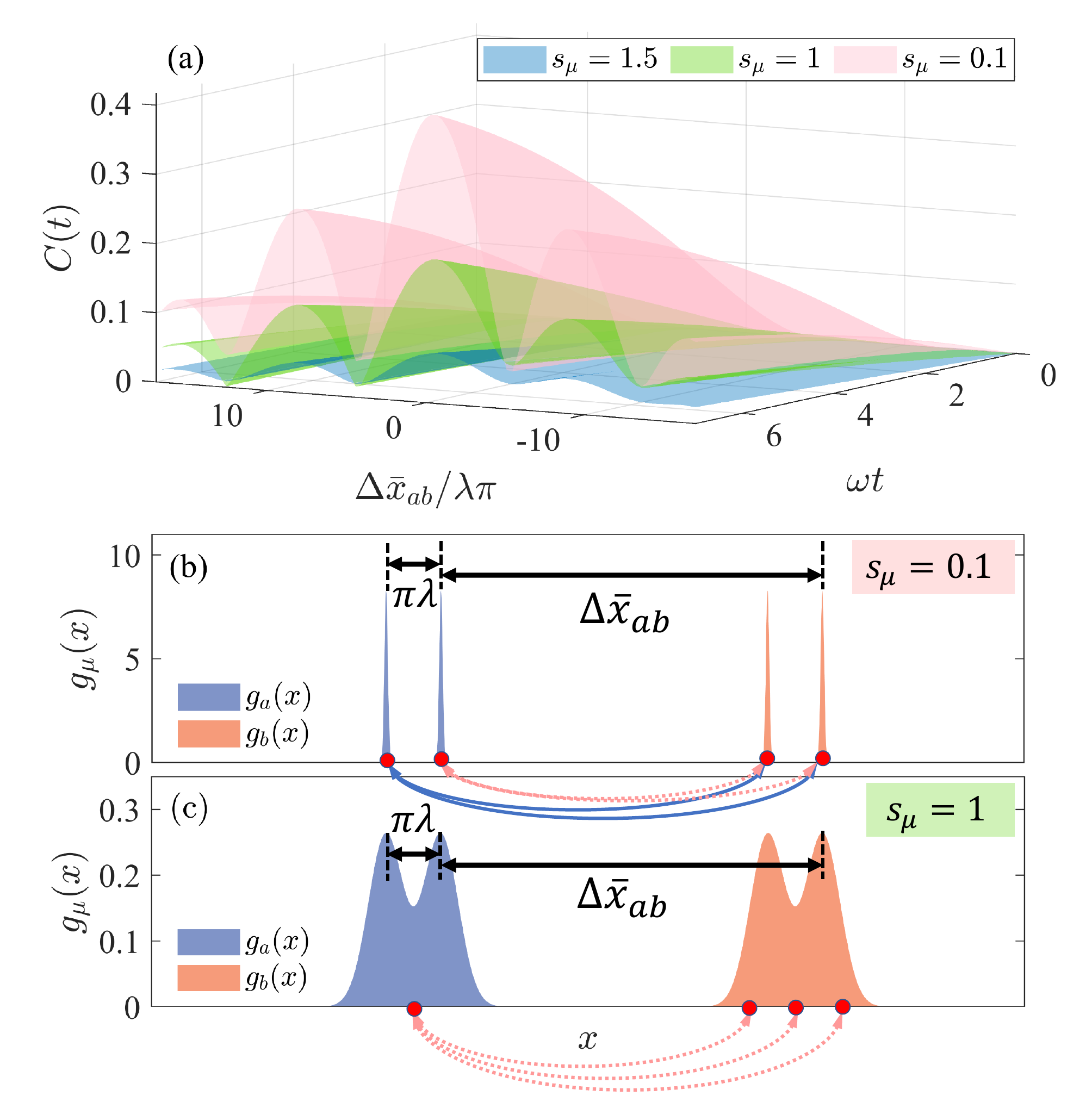} \caption{The weakening of interference effects as a consequence of the broadening
of the continuous coupling. (a) Entanglement generation for different
spatial distributions of the continuous coupling $g_{\mu}(x)$ given
in Eq.~(\ref{eq:gx_2peak}). (b) and (c) are the distribution functions
$g_{a}(x)$ and $g_{b}(x)$ for the case $s_{\mu}=0.1$ and $s_{\mu}=1$
(assuming $s_{a}=s_{b}$), respectively.}
\label{fig:5}
\end{figure}

As shown in Fig.~\ref{fig:5}(b) and (c), each giant
atom couples to the waveguide in two spatially separated regions centered
at $\bar{x}_{\mu1}$ and $\bar{x}_{\mu2}$. For both giant atoms,
we now fix the inter-peak distance of $g_{\mu}(x)$ as $\bar{x}_{a2}-\bar{x}_{a1}=\bar{x}_{b2}-\bar{x}_{b1}=\pi\lambda$.
Then, the only crucial parameter becomes the distance
\begin{equation}
\Delta\bar{x}_{ab}=\bar{x}_{b1}-\bar{x}_{a1}=\bar{x}_{b2}-\bar{x}_{a2},
\end{equation}
which describes spatial separation between the centers of the relevant
coupling regions corresponding to the two giant atoms ``$a$'' and
``$b$'' {[}see Fig.~\ref{fig:5}(b) and (c){]}. The parameter
$\Delta\bar{x}_{ab}$ determines the phase accumulated when a photon
travels from one giant atom to the other one. In the strongly-localized
distribution case $s_{\mu}=0.1$, as shown in Fig.~\ref{fig:5}(b),
there is a strong interference because the phase difference accumulated
is almost a constant. This is reflected in the entanglement generation
shown in Fig.~\ref{fig:5}(a). The pink surface (for $s_{\mu}=0.1$)
exhibits a pattern analogous to interference fringes when $\Delta\bar{x}_{ab}$
is varied.

In contrast, as $s_{\mu}$ increases {[}e.g., $s_{\mu}=1$
in Fig.~\ref{fig:5}(c){]}, the double-peak distribution of $g_{\mu}(x)$
broadens significantly. Each giant atom couples to the waveguide in
a much broader region. As illustrated in Fig.~\ref{fig:5}(c), there
are many possible pathways for the photon to be emitted and absorbed
by the two giant atoms, as a result, the phase accumulated by a photon
traveling between the two atoms is no longer nearly constant. The
breaking of constant phase different condition eventually weaken the
interference pattern as shown in the green surface in Fig.~\ref{fig:5}(a).

When $s_{\mu}$ is further increased to $s_{\mu}=1.5$,
the corresponding blue surface in Fig.~\ref{fig:5}(a) appears nearly
flat, with only faint variations in entanglement generation. The periodic
interference pattern has been erased by the broadened continuous coupling.
This trend confirms that the broadening of continuous coupling directly
suppresses the interference effects inherent to giant-atom systems.

\section{Multiple excitations}

\label{double} In Sec.~\ref{sec:Continuous}, we
mainly focus on the first gap that mentioned in Sec.~\ref{sec:Continuous}.
In this section, we will try to fill the second gap and we extend
the analysis beyond the single excitation subspace to investigate
the dynamical properties within two- and multiple-excitations. We
will start from the increased complexity of the conventional methods.
\subsection{Complexity of the conventional finite-excitation
approach}\label{subsec:6.1}
The conventional dressed state approach has been
reviewed in Sec.~\ref{sec:SingleExcitation}. Now, we go beyond the
single excitation subspace in Eq.~(\ref{eq:SingleExcitationState})
to assume two excitations. In the two-excitation subspace, the state
vector can be expressed as
\begin{equation}
\begin{aligned}|\psi^{(2)}(t)\rangle= & C_{ab}(t)\sigma_{a}^{+}\sigma_{b}^{+}|G\rangle+\sum_{\mu=a,b}\sum_{k}B_{\mu,k}(t)\sigma_{\mu}^{+}c_{k}^{\dagger}|G\rangle\\
 & +\sum_{k\neq k^{\prime}}A_{k,k^{\prime}}(t)c_{k}^{\dagger}c_{k^{\prime}}^{\dagger}|G\rangle,
\end{aligned}
\end{equation}
Consequently, the dynamical equation.~(\ref{eq:dressstate}) must
be reformulated into the following more intricate form,
\begin{equation}
\begin{aligned}\dot{C}_{ab}(t)= & -i\sum_{\mu=a,b}\sum_{k}G_{\mu k}e^{i(\omega_{k}-\omega_{\mu})t}B_{\mu,k}(t),\\
\dot{B}_{\mu,k}(t)= & -iG_{\nu k}^{*}e^{-i(\omega_{k}-\omega_{\nu})t}C_{ab}(t)\\
 & -i\sum_{k^{\prime}}G_{\mu k}e^{i(\omega_{q}-\omega_{\mu})t}A_{k,k^{\prime}}(t),\ (\mu\ne\nu)\\
\dot{A}_{k,k^{\prime}}(t)= & -i\sum_{\mu=a,b}G_{\mu k}^{*}e^{-i(\omega_{k}-\omega_{\mu})t}B_{\mu,k^{\prime}}(t)\\
 & -i\sum_{\mu=a,b}G_{\mu k^{\prime}}^{*}e^{-i(\omega_{k^{\prime}}-\omega_{\mu})t}B_{\mu,k}(t),
\end{aligned}
\end{equation}
where $C_{ab}(t)$ describes the double-atom excited state, $B_{\mu,k}(t)\ (\mu=a,b)$
describes the hybrid atom-waveguide excited states, and $A_{k,k^{\prime}}(t)$
corresponds to the double-waveguide excited state.

It is evident that employing the dressed state approach
in the two-excitation regime leads to substantial complexity, posing
significant challenges for both analytical derivation and numerical
simulation. To consider more excitations, the complexity of analytical
and numerical computation will be sharply increased. In contrast,
the SSE approach in Sec.~\ref{sec:SSEA} is independent of the specific
form of the initial state as well as the total excitations. Compared
to the single-excitation case, numerical simulations for two and more
excitations only require a change of the initial state vector. Thus,
the SSE approach demonstrates a clear advantage for addressing such
problems and provides a valuable tool for research in this field.

\subsection{Double excitation state $|ee\rangle$}

\begin{figure}[htbp]
\centering \includegraphics[width=1\columnwidth]{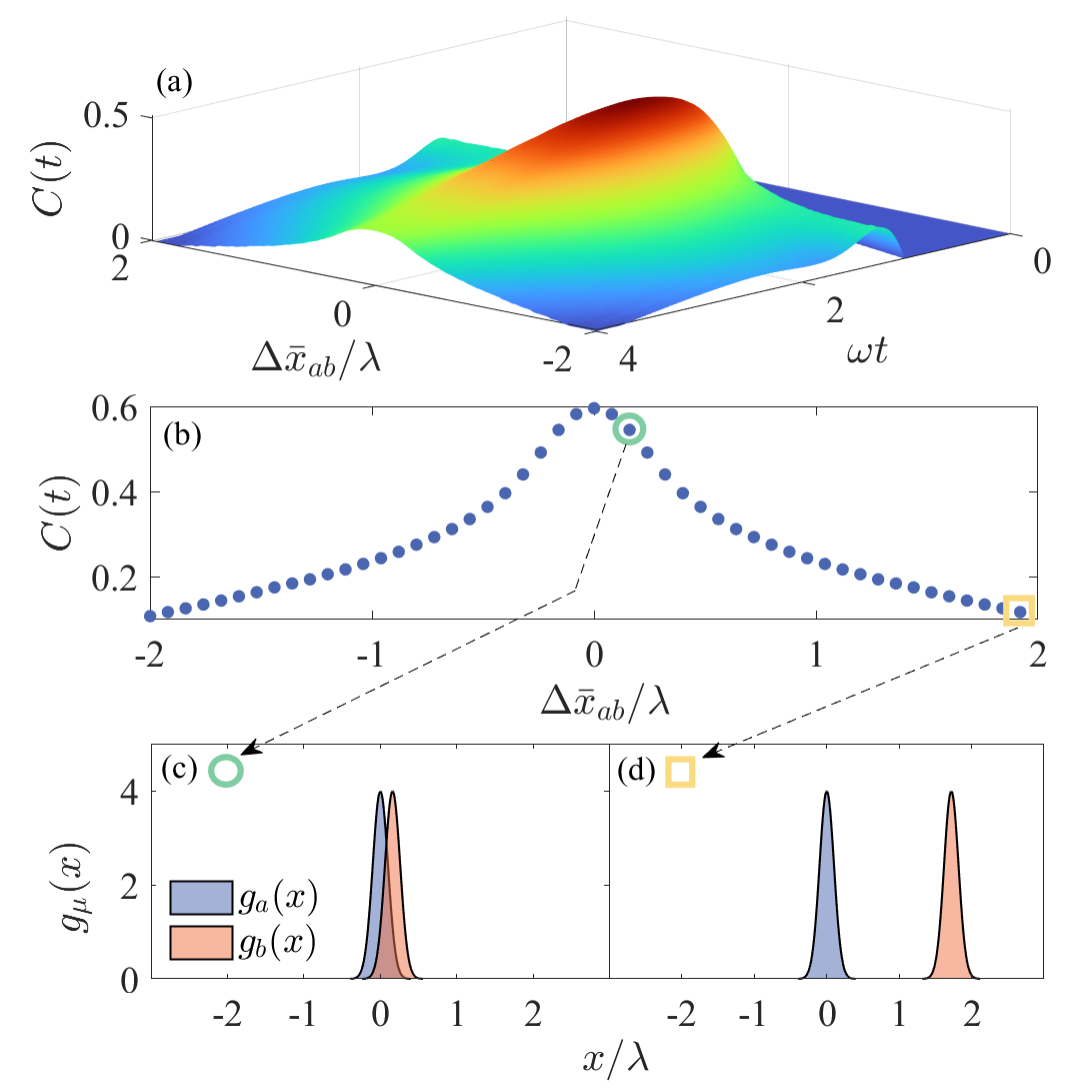} \caption{Entanglement generation from the initial state $|ee\rangle$ for strongly
localized coupling ($s_{a}=s_{b}=0.1$). (a) 3D plot of concurrence
$C(t)$ versus time $t$ and the distance between two centers of the
coupling $\Delta\bar{x}_{ab}=\bar{x}_{b}-\bar{x}_{a}$. (b) 2D cross-section
of $C(t)$ at $\omega t=1.8$. (c,d) Coupling distributions for: (c)
$\bar{x}_{a}=0$, $\bar{x}_{b}=0.16\lambda$; (d) $\bar{x}_{a}=0$,
$\bar{x}_{b}=1.92\lambda$.}
\label{fig:6} 
\end{figure}

\begin{figure}[htbp]
\centering \includegraphics[width=1\columnwidth]{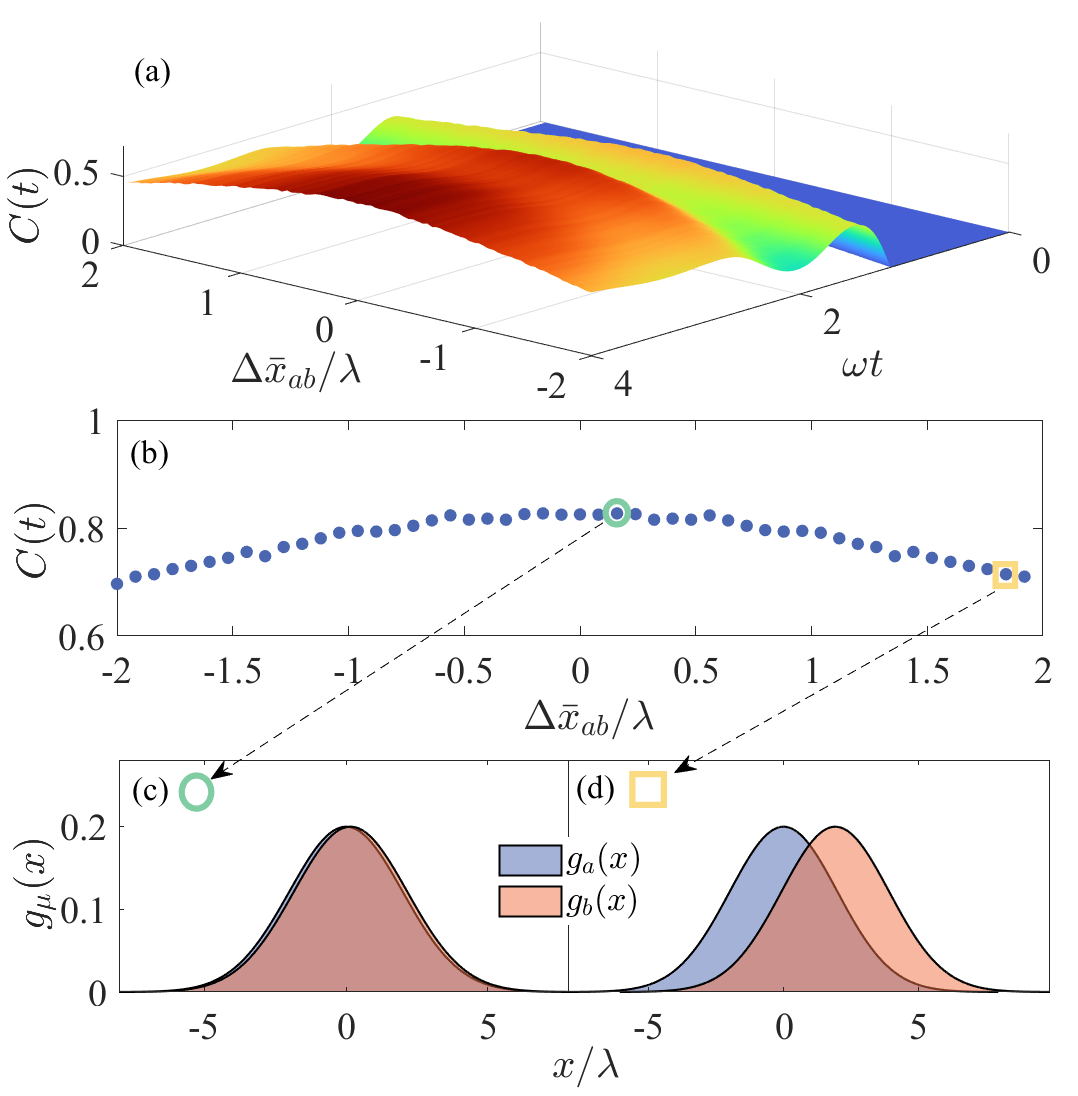} \caption{Entanglement generation from the initial state $|ee\rangle$ for strongly
localized coupling ($s_{a}=s_{b}=2$). (a) 3D plot of concurrence
$C(t)$ versus time $t$ and $\Delta\bar{x}_{ab}$. (b) 2D cross-section
at $\omega t=1.8$ for $C(t)$. (c,d) Coupling distribution for: (c)
$\bar{x}_{a}=0$, $\bar{x}_{b}=0.16\lambda$; (d) $\bar{x}_{a}=0$,
$\bar{x}_{b}=1.92\lambda$.}
\label{fig:7} 
\end{figure}

To demonstrate the advantage of the SSE approach, we begin with the
entanglement generation process of a double excitation state $|ee\rangle$
state. Here, we assume the single peak Gaussian distribution for the
continuous coupling 
\begin{equation}
g_{\mu}(x)=\frac{1}{s_{\mu}\sqrt{2\pi}}\exp\left[-\frac{(x-\bar{x}_{\mu})^{2}}{2s_{\mu}^{2}}\right].\quad(\mu=a,b)
\end{equation}
Comparing the numerical results in Fig.~\ref{fig:6} and Fig.~\ref{fig:7},
we have two observations. First, the entanglement $C(t)$ in the localized
regime in Fig.~\ref{fig:6} has a much stronger position dependent
than the case in Fig.~\ref{fig:7}. The cross-section of entanglement
$C(t)$ in Fig.~\ref{fig:6}(b) is very sensitive to the distance
$\Delta\bar{x}_{ab}=\bar{x}_{b}-\bar{x}_{a}$, while $C(t)$ in Fig.~\ref{fig:7}(b)
is insensitive to $\Delta\bar{x}_{ab}=\bar{x}_{b}-\bar{x}_{a}$. Second,
the overall entanglement generation in the localized regime {[}Fig.~\ref{fig:6}(a){]}
is weaker than the delocalized regime {[}Fig.~\ref{fig:7}(a){]}.
This is reflected either from the maximum entanglement or from the
average entanglement. The peak in Fig.~\ref{fig:6}(b) is less than
$0.6$, while the entanglement $C(t)$ in Fig.~\ref{fig:7}(b) is
always larger than $0.6$ for any $\Delta\bar{x}_{ab}$, the peak
is even larger than $0.8$.

Both of the above two observations can be explained from the physical
picture of photon emission and absorption. First, in the localized
regime in Fig.~\ref{fig:6}, photons can be only emitted almost at
a single point, and then absorbed also almost at a single point. The
phase accumulation in the propagation process is close to a constant,
which is sensitive to the distance $\Delta\bar{x}_{ab}=\bar{x}_{b}-\bar{x}_{a}$.
Thus, a strong $\Delta\bar{x}_{ab}$ dependence is observed in Fig.~\ref{fig:6}.
As a comparison, in the delocalized regime in Fig.~\ref{fig:7},
photons can be emitted and absorbed at any points in the range of
$g_{a}(x)$ and $g_{b}(x)$. There are a large amount of trajectories
corresponding to different phase accumulations. The influence of the
distance $\Delta\bar{x}_{ab}$ is smoothed out through averaging over
all possible trajectories. Second, the delocalized distribution in
Fig.~\ref{fig:7} also provides more possibilities to build connections
between two giant atoms through photon emission and absorption. Since
there is not direct coupling between two giant atoms in the Hamiltonian~(\ref{eq:Htot}),
two giant atoms can be only indirectly coupled through the waveguide.
More photon emission and absorption process means stronger indirect
coupling, leading to stronger entanglement generation.

\subsection{Thermal and squeezed initial states of the waveguide}

\label{ThermalSqueezed}

The SSE approach naturally extends to higher-excitation subspaces,
providing a powerful tool to explore multi-photon initial state of
the waveguide. For example, as we have discussed in Sec.~\ref{sec:Introduction},
the the continuous spectrum of waveguide modes naturally involves
the thermal initial state at finite temperatures. In Appendix~\ref{sec:AppThermal},
we demonstrate that the SSE approach is compatible with thermal initial
states. To be specific, one can introduce a set of fictitious modes
with negative frequencies which has no interactions to both the giant
atoms and the waveguide modes. By performing a Bogoliubov transformation,
the thermal initial state of the waveguide is transformed into combined
vacuum state for waveguide modes plus fictitious modes. Since the
fictitious modes are isolated from other parts, it has no influence
on the original dynamics. Thus, solving the model with the fictitious
mode is equivalent to solving the original model with thermal initial
states. The detailed derivation and discussion are presented in Appendix~\ref{sec:AppThermal}.

The SSE approach also applicable to other nonclassical initial states,
such as squeezed states (Appendix~\ref{sec:AppSqueeze}). This enables
the investigation of quantum dynamics in highly excited regimes and
providing new perspectives on nonlinear quantum optical phenomena
\cite{Yin2025}. Although the detailed discussion can be left in a
future study, we notice that the cross-correlation functions are enhanced
{[}see Eq.~(\ref{eq:sqnoise}) and Eq.~(\ref{eq:sqCF}){]}. This
implies the possibility of exponential enhancement \cite{Kam2024,Qin2022,Qin2018}
of cross-correlation function. From physical perspective, it means
exponential enhancement of indirect interactions between two giant
atoms. Therefore, our approach provide a theoretical tool for the
future study for enhanced interactions between giant atoms via squeezed
states in waveguide. 

Furthermore, this tool can be extended to investigate
the role of the spatial structure of squeezing \cite{Gutierrez2023PRR},
which is anticipated to yield novel control over photon-mediated interactions
in giant atom systems. This is also discussed in Appendix~\ref{sec:AppSqueeze}.

\section{Conclusion}\label{sec:Conclusion}

This work establishes a systematic theoretical framework
for investigating the dynamics of giant atoms coupled to a waveguide,
which is applicable to continuous coupling and multiple-excitation
scenarios and is further applied to study the entanglement dynamics
of two giant atoms coupled to a single waveguide. To summarize, our
work features the following key highlights.

(a) In the SSE approach, the auto- and cross-correlation
functions offer a distinct perspective for quantifying the impact
of time-delay effects.

(b) Continuous coupling yields distinct physical
insights compared to the discrete coupling model widely studied in
previous studies. Unlike discrete coupling where fixed optical paths
enforce a constant phase difference, continuous coupling occurs over
a spatial region, breaking the constant phase difference condition
and thereby weakening interference-induced phenomena. This addresses
Gap 1 identified in Sec.~\ref{sec:Introduction}.

(c) The SSE approach naturally handles multiple excitations.
In traditional ``dressed state approach'', the complexity of the
dynamical equations increases drastically with the number of excitations.
In contrast, in SSE approach, the dynamical equation remains unchanged
as the number of excitations grows, greatly reducing the difficulty
of numerical computations. Besides, we demonstrate that the SSE approach
is compatible with both thermal and squeezed initial states, which
opens a new window for future study on the introduction of squeezing
in waveguides. This partially fills Gap 2 as mentioned in Sec.~\ref{sec:Introduction}.

(d) An additional contribution of this work is the
incorporation of two types of noise calculations into the SSE framework,
thereby extending the original SSE theoretical formalism~\cite{PhysRevA.58.1699,PhysRevLett.82.1801,PhysRevA.60.91}.

We anticipate that the developed SSE framework will
serve as a versatile platform for future investigations into the dynamics
of giant atom-waveguide systems, particularly for exploring complex
scenarios involving continuous coupling and multiple excitations.
\begin{acknowledgments}
We thank the support from the National Natural Science Foundation
of China under Grant No. 62471143, the Key Program of National Natural
Science Foundation of Fujian Province under Grant No. 2024J02008,
the project from Fuzhou University under Grant No. JG2020001-2, and
the Natural Science Foundation of Fujian Province under Grant No.
2022J01548.
\end{acknowledgments}

\appendix

\section{Correlation function }\label{sec:AppCF}

There are two equivalent ways to define the correlation functions.
The first one is presented in the main text, i.e.,
\begin{equation}
\alpha_{\mu\mu}(t,s)=\sum_{k}|G_{\mu k}|^{2}e^{-i\omega_{k}(t-s)}.
\end{equation}
In this definition, $\alpha_{\mu\mu}$ (as an example, the other ones
are similar) is interpreted as a Fourier transform of $|G_{\mu k}|^{2}$.
The second definition is based on the collective coupling operator
defined as $C_{\mu}(t)\equiv\sum_{k}G_{\mu k}c_{k}^{\dag}e^{i\omega_{k}t}$.
Given this operator, the interaction Hamiltonian can be written as
\begin{align}
H_{{\rm tot}}^{I}(t) & =H_{{\rm A}}+\sum_{\mu=a,b}\left(\sum_{k}G_{\mu k}\sigma_{\mu}^{-}c_{k}^{\dag}e^{i\omega t}+{\rm H.c.}\right)\nonumber \\
 & =H_{{\rm A}}+\sum_{\mu=a,b}\left(\sigma_{\mu}^{-}C_{\mu}^{\dag}+{\rm H.c.}\right).
\end{align}
Then, the correlation function can be expressed as the expectation
value of operator $C_{\mu}(t)C_{\mu}^{\dag}(s)$ in the vacuum state,
i.e.,
\begin{equation}
\alpha_{\mu\mu}(t,s)=\langle0|C_{\mu}(t)C_{\mu}^{\dag}(s)|0\rangle.
\end{equation}
One can check these two definition are equivalent \cite{Zhao2025OE}.
The second definition reveals from another perspective that the correlation
function reflects the nature of the coupling between the giant atoms
and the waveguide.

\section{Master equation}\label{sec:AppMEQ}

The density operator can be reconstructed by averaging all the possible
stochastic quantum trajectories as shown in Eq.~(\ref{eq:Mean}).
Define $P_{t}\equiv|\psi(t,z^{*})\rangle\langle\psi(t,z)|$ as the
stochastic density operator, then take the time derivative of $\rho=\mathcal{M}\{P_{t}\}$,
yielding 

\begin{equation}
\begin{aligned}\frac{d}{dt}\rho= & \frac{d}{dt}\mathcal{M}\{P_{t}\}\\
= & \mathcal{M}\left\{ \left[\frac{d}{dt}|\psi(t,z^{*})\rangle\right]\langle\psi(t,z)|\right\} \\
 & +\mathcal{M}\left\{ |\psi(t,z^{*})\rangle\left[\frac{d}{dt}\langle\psi(t,z)|\right]\right\} \\
= & \mathcal{M}\left\{ H_{{\rm eff}}P_{t}\right\} +\mathcal{M}\left\{ P_{t}H_{{\rm eff}}\right\} .
\end{aligned}
\label{eq:drho}
\end{equation}
In the above equation, we employ the SSE in Eq.~(\ref{eq:NMQSD})
to substitute $\frac{d}{dt}|\psi(t,z^{*})\rangle$ with $H_{{\rm eff}}|\psi(t,z^{*})\rangle$.
It is straightforward to show that $\mathcal{M}\{-iH_{\text{A}}P_{t}\}=-iH_{\text{A}}\mathcal{M}\{P_{t}\}=-iH_{\text{A}}\rho$,
since $H_{\text{A}}$ is independent of the noise variables $z$ or
$z^{*}$. However, computing $\mathcal{M}\{\sigma_{\mu}^{-}z_{t}^{*}P_{t}\}$
is no longer a trivial task. To evaluate this term, we will invoke
the Novikov theorem as 
\begin{equation}
\mathcal{M}\{P_{t}z_{\mu t}\}=\mathcal{M}\{\bar{O}_{\mu}P_{t}\},\label{eq:NovikovBoson1}
\end{equation}
\begin{equation}
\mathcal{M}\{z_{\mu t}^{*}P_{t}\}=\mathcal{M}\{P_{t}\bar{O}_{\mu}^{\dagger}\}.\label{eq:NovikovBoson2}
\end{equation}
Here is a brief proof of the Novikov theorem, which is similar to
the proof given in Refs. \cite{PhysRevA.86.032116,PhysRevA.60.91}.
By definition, 
\begin{widetext}
\begin{eqnarray}
\mathcal{M}\{P_{t}z_{\mu t}\} & = & {\displaystyle \int}{\displaystyle \prod\nolimits_{k}}dz_{k}^{\ast}dz_{k}\exp\left(-\sum_{k}z_{k}^{\ast}z_{k}\right)P_{t}\left(i\sum_{j}G_{j}^{*}e^{-i\omega_{j}t}z_{\mu j}\right)\nonumber \\
 & = & -i\sum_{j}G_{j}^{*}e^{-i\omega_{j}t}{\displaystyle \int}{\displaystyle \prod\nolimits_{k}}dz_{k}^{\ast}dz_{k}P_{t}\frac{\partial}{\partial z_{\mu j}^{\ast}}\left[\exp\left(-\sum_{k}z_{k}^{\ast}z_{k}\right)\right].
\end{eqnarray}
Integrating by parts and noting that in polar coordinates, $[re^{-|r|^{2}}]_{0}^{\infty}=0$,
one obtain 
\begin{eqnarray}
\mathcal{M}\{P_{t}z_{\mu t}\} & = & i\sum_{j}G_{\mu j}^{*}e^{-i\omega_{j}t}z_{\mu j}{\displaystyle \int}{\displaystyle \prod\nolimits_{k}}dz_{k}^{\ast}dz_{k}\frac{\partial}{\partial z_{\mu j}^{\ast}}P_{t}\exp\left(-\sum_{k}z_{k}^{\ast}z_{k}\right)\nonumber \\
 & = & i\sum_{j}G_{\mu j}^{*}e^{-i\omega_{j}t}z_{\mu j}{\displaystyle \int}{\displaystyle \prod\nolimits_{k}}dz_{k}^{\ast}dz_{k}\left(\int ds\frac{\partial z_{\mu s}^{\ast}}{\partial z_{\mu j}^{\ast}}\frac{\delta}{\delta z_{\mu s}^{\ast}}\right)P_{t}\exp\left(-\sum_{k}z_{k}^{\ast}z_{k}\right)\nonumber \\
 & = & \int ds\sum_{j}\left\vert G_{\mu j}\right\vert ^{2}e^{-i\omega_{j}(t-s)}{\displaystyle \int}{\displaystyle \prod\nolimits_{k}}dz_{k}^{\ast}dz_{k}\exp\left(-\sum_{k}z_{k}^{\ast}z_{k}\right)\frac{\delta}{\delta z_{\mu s}^{\ast}}P_{t}\nonumber \\
 & = & \mathcal{M}\{\bar{O}_{\mu}P_{t}\}.
\end{eqnarray}
Similarly, we can also prove 
\begin{equation}
\mathcal{M}\{z_{\mu t}^{*}P_{t}\}=\mathcal{M}\{P_{t}\bar{O}_{\mu}^{\dagger}\}.
\end{equation}
\end{widetext}

Applying the Novikov theorem (Eqs. \ref{eq:NovikovBoson1} and \ref{eq:NovikovBoson2})
to Eq. \ref{eq:drho}, we arrive at the final master equation as 
\begin{equation}
\begin{aligned}\frac{\partial}{\partial t}\rho= & -i[H_{\text{A}},\rho]+\sum_{\mu=a,b}\left(\left[\sigma_{\mu}^{-},\mathcal{M}[P_{t}\bar{O}_{\mu}^{\dagger}]\right]+{\rm H.c.}\right).\label{gaqsd}\end{aligned}
\end{equation}
If $O_{\mu}$ operator is independent of noise variable
$z$, the master equation will reduce to
\begin{equation}
\frac{\partial}{\partial t}\rho=-i[H_{\text{A}},\rho]+\sum_{\mu=a,b}\left([\sigma_{\mu}^{-},\rho\bar{O}_{\mu}^{\dagger}]+{\rm H.c.}\right).
\end{equation}
This is the master equation presented in Eq.~(\ref{eq:MEQ})
in the main text.

In Eq.~(\ref{eq:Oab}), the solution of $O_{\mu}$
operator is an approximate solution. The exact solution of $O_{\mu}$
contains five terms with the fifth term associated with the operator
$O_{\mu5}=\sigma_{\mu}^{-}\sigma_{\nu}^{-}$ to be $z_{\mu t}^{*}$-dependent
\cite{PhysRevA.84.032101}. Here, we take the noise free $O$ operator
$O_{\mu}^{(0)}(t,s)=\sum_{j=1}^{4}p_{\mu j}(t,s)O_{\mu j}$ as an
approximate solution for computational efficiency, the calculation
will be hugely reduced, since $\mathcal{M}[P_{t}\bar{O}_{\mu}^{\dagger}]=\rho\bar{O}_{\mu}^{\dagger}$.
Meanwhile, the accuracy of this approximation is remarkable in many
cases~\cite{Xu_2014}. In Ref.~\cite{Xu_2014}, the authors show
that the correction from the fifth term generally contributes only
to fourth-order (or higher) effects in the coupling strengths. When
$G_{\mu k}$ is smaller than $\omega_{\mu}$, the noise term becomes
negligible. Certainly, if necessary, one can derive the exact solution
by following Ref.~\cite{PhysRevA.84.032101} in a straightforward
manor. If the noise-dependent term $O_{\mu5}$ is taken into consideration,
one can still derive a master equation by following the procedure
in Ref.~\cite{Chen2014PRA}.

\section{Entanglement generation for continuous coupling with Gaussian distribution}\label{sec:AppC}

\begin{figure}[tb]
\centering \includegraphics[width=1\columnwidth]{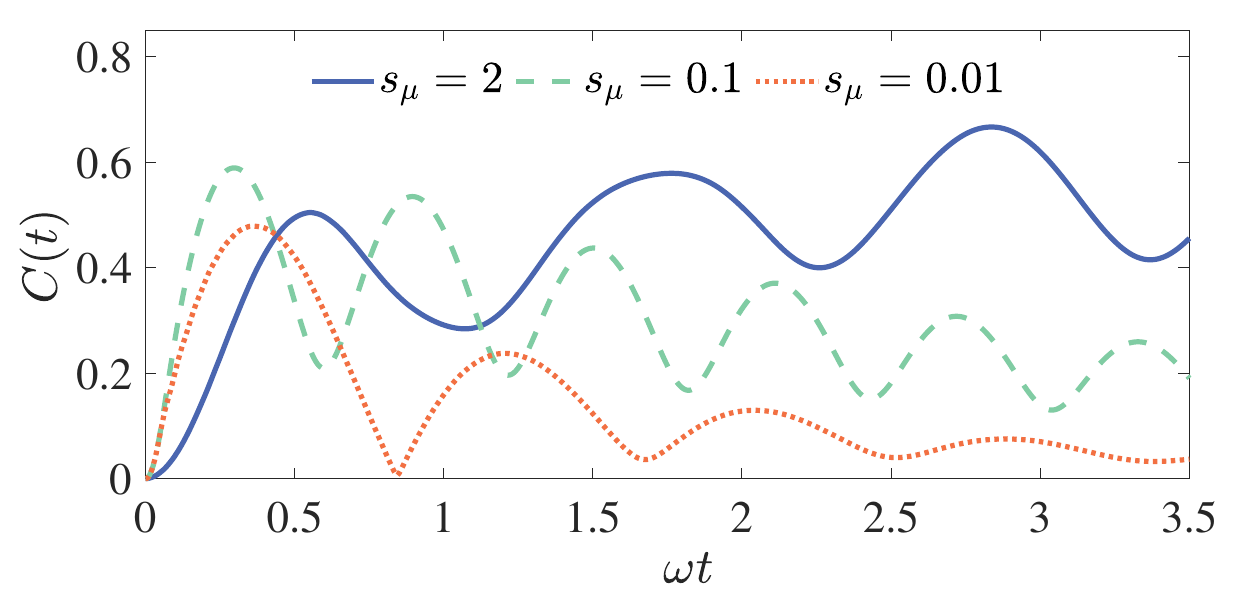} \caption{Concurrence $C(t)$ for different Gaussian coupling distributions
$s_{\mu}=0.01$ (orange dotted), $s_{\mu}=0.1$ (green dashed), and
$s_{\mu}=2$ (blue solid).}
\label{fig:8}
\end{figure}

In Sec.~\ref{subsec:Gaussian}, we analyzed the continuous coupling
with spatial Gaussian distribution from the perspective of correlation
functions. In this section, we will investigate the influence of Gaussian
broadening on entanglement dynamics from the viewpoint of entanglement
generation. 

The numerical results in Fig.~\ref{fig:8} presents
$C(t)$ evolution for three $s_{\mu}$ values. For $s_{\mu}=0.01$
(extremely localized, close to the case of small atoms), coupling
is nearly point-like, leading to weak entanglement with rapid decay.
For $s_{\mu}=0.1$ (moderately localized), broader coupling introduces
more photon paths, enhancing entanglement strength and stability with
obvious revival. For $s_{\mu}=2$ (strongly delocalized), photons
are emitted/absorbed arbitrarily across the wide region, resulting
in the highest peak concurrence and robust long-term stability.

The trend in Fig.~\ref{fig:8} is that larger $s_{\mu}$
(stronger delocalization) improves entanglement performance. As the
distribution width $s_{\mu}$~increases, it corresponds
to the transition from small atoms to giant atoms. Compared to the
case of small atoms illustrated in Fig.~\ref{fig:1}(a), giant atoms
offer far more diverse photon emission and absorption pathways. This
creates additional opportunities for building correlations between
the two atoms, thus facilitating the generation of entanglement.

\section{The influence of center position}\label{appendix:CenterPosition}

\begin{figure}[htbp]
\centering \includegraphics[width=1\columnwidth]{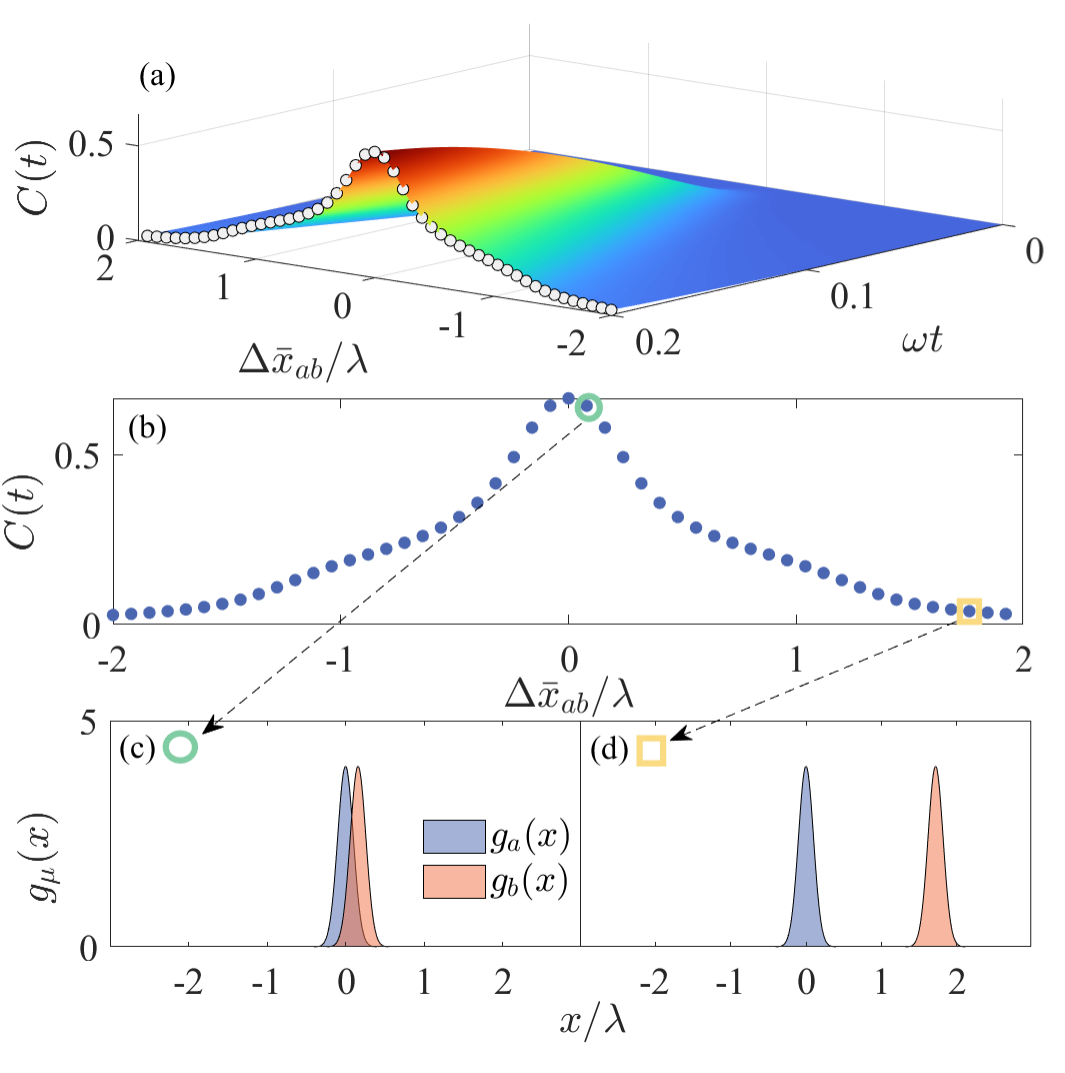} \caption{Entanglement generation in localized regime ($s_{\mu}=0.1$). (a)
Concurrence $C(t)$ as a function of $\Delta\bar{x}_{ab}$ and $\omega t$.
(b) Cross-section at $\omega t=0.2$. (c-d) Examples of coupling distributions
at $\Delta\bar{x}_{ab}=0.08\lambda$ and $\Delta\bar{x}_{ab}=1.92\lambda$.}
\label{fig:9} 
\end{figure}

In this section, we show the influence of the center position of the
coupling distribution. In the Gaussian distribution in Eq.~(\ref{eq:GaussianDis}),
$\bar{x}_{a}$ and $\bar{x}_{b}$ determine the center of the coupling
profile for two giant atoms. From the numerical results in Fig.~\ref{fig:9},
in the localized coupling regime ($s_{\mu}=0.1$), the concurrence
$C(t)$ exhibits strong dependence on the center positions $\bar{x}_{\mu}$
of the coupling distribution {[}Fig.~\ref{fig:9}(a){]}, arising
from photon-mediated interactions between giant atoms. The cross-section
of concurrence at $\omega t=0.2$ {[}Fig.~\ref{fig:9}(b){]} clearly
demonstrates spatial modulation. Actually, the residue entanglement
depends on $\Delta\bar{x}_{ab}=\bar{x}_{a}-\bar{x}_{b}$ since $\Delta\bar{x}_{ab}$
corresponds to the phase accumulation between the photon emission
and reabsorption.

\begin{figure}[htbp]
\centering \includegraphics[width=1\columnwidth]{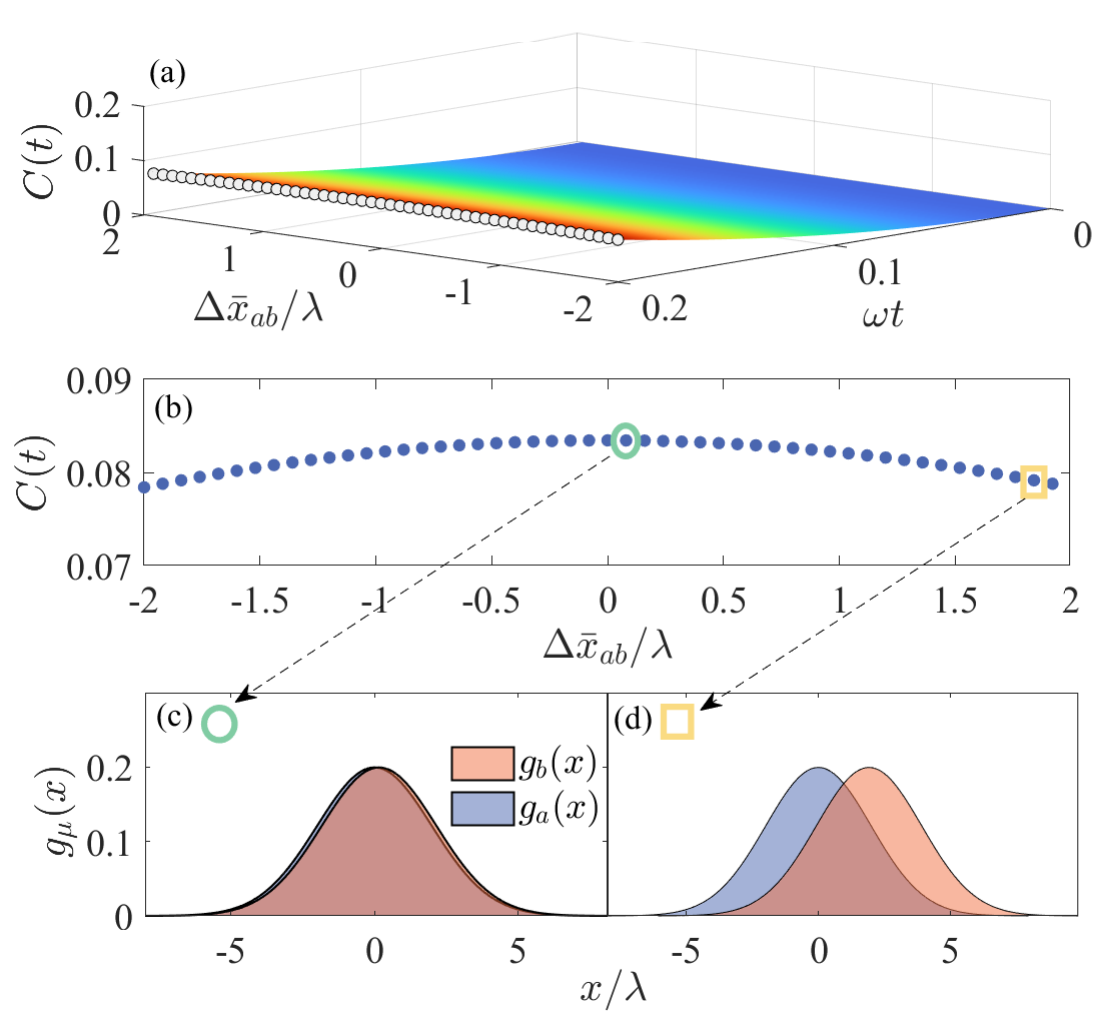} \caption{Entanglement generation in delocalized regime ($s_{\mu}=4$). (a)
Weak $\Delta\bar{x}_{ab}$-dependence of $C(t)$ evolution. (b) Cross-section
of $C(t)$ at $\omega t=0.2$ . (c-d) Examples of coupling distributions
at $\Delta\bar{x}_{ab}=0.08\lambda$ and $\Delta\bar{x}_{ab}=1.92\lambda$.}
\label{fig:10} 
\end{figure}

In the delocalized coupling regime ($s_{\mu}=4$), the system exhibits
markedly different behavior: Entanglement becomes highly insensitive
to the center position of coupling distribution {[}Fig.~\ref{fig:10}(a){]}.
The two-dimensional concurrence profile at $\omega t=0.2$ versus
$\Delta\bar{x}_{ab}$ {[}Fig.~\ref{fig:10}(b){]} shows minimal fluctuations,
demonstrating exceptional robustness. This stability originates from
extended coupling distributions creating multiple interaction channels
between photons. Taking the photon emission-reabsorption by different
atoms as an example. The photon can be emitted in any position in
the region of $g_{a}(x)$, and then reabsorbed in any position in
the region of $g_{b}(x)$. The phase accumulated for each possible
emission-reabsorption path can be quite different, and the averaging
of all trajectories has smoothed out the impact caused by the center
position.

\section{Thermal initial states in waveguide}\label{sec:AppThermal}

In the case of the finite-temperature waveguide, we can transform
the finite temperature case into an effective zero temperature model
by introducing a fictitious bath~\cite{PhysRevA.69.062107}. The
initial thermal state of the waveguide can be written as 
\begin{equation}
\rho_{\text{W}}(0)=\frac{e^{-\beta H_{\text{W}}}}{Z},
\end{equation}
where $Z={\rm tr}[e^{-\beta H_{\text{W}}}]$ is the partition function
with $\beta=\frac{1}{k_{\text{B}}T}$. The occupation number for mode
$k$ should be 
\begin{equation}
\langle c_{k}^{\dagger}c_{k}\rangle=\frac{1}{e^{\beta\hbar\omega_{k}}-1}.\label{bk}
\end{equation}
This is the well known Bose-Einstein distribution. By introducing
a fictitious mods with negative frequencies $H_{\text{B}}=-\sum_{k}\omega_{k}b_{k}^{\dagger}b_{k}$,
the finite temperature problem can be mapped into a zero temperature
problem. After adding a fictitious bath, the total Hamiltonian become
\begin{equation}
\begin{aligned}H_{\text{tot}}= & H_{\text{A}}+\sum_{k}\omega_{k}c_{k}^{\dagger}c_{k}+\sum_{\mu=a,b}\sum_{k}(g_{\mu k}c_{k}^{\dagger}\sigma_{\mu}^{-}+H.c.)\\
 & -\sum_{k}\omega_{k}b_{k}^{\dagger}b_{k}.
\end{aligned}
\end{equation}
It should be noted that the fictitious bath $H_{\text{B}}$ has no
direct interaction with either the giant atoms or the waveguide. Therefore
it has no impact on the dynamics of the original Hamiltonian (system
plus the original waveguide). Namely, solving this Hamiltonian is
equivalent to solving the original Hamiltonian.

After a Bogoliubov transformation 
\begin{align}
c_{k} & =\sqrt{\bar{n}_{k}+1}d_{k}+\sqrt{\bar{n}_{k}}e_{k}^{\dagger},\\
b_{k} & =\sqrt{\bar{n}_{k}+1}e_{k}+\sqrt{\bar{n}_{k}}d_{k}^{\dagger},
\end{align}
the Hamiltonian becomes 
\begin{equation}
\begin{aligned}H_{\text{tot}}^{\prime}= & H_{\text{A}}+\sum_{k}\omega_{k}d_{k}^{\dagger}d_{k}+\sum_{\mu,k}\sqrt{\bar{n}_{k}+1}(g_{\mu k}\sigma_{\mu}^{-}d_{k}^{\dagger}+H.c.)\\
 & -\sum_{k}\omega_{k}e_{k}^{\dagger}e_{k}+\sum_{\mu,k}\sqrt{\bar{n}_{k}}(g_{\mu k}\sigma_{\mu}^{-}e_{k}+H.c.).
\end{aligned}
\label{eq:HT}
\end{equation}
It is easy to check the vacuum state for the system $|0\rangle=|0\rangle_{d}\otimes|0\rangle_{e}$
satisfies $d_{k}|0\rangle=0$, and $e_{k}|0\rangle=0$, and it keeps
the original Bose-Einstein distribution at temperature $T$ in Eq.~(\ref{bk}).
Therefore, solving the transformed Hamiltonian~(\ref{eq:HT}) with
the initial state $|0\rangle_{d}\otimes|0\rangle_{e}$ is equivalent
to solving the original Hamiltonian with the thermal initial state
$\rho_{\text{W}}(0)=e^{-\beta H_{\text{W}}}/Z$. Thus, the finite
temperature problem is transformed into a zero temperature problem.
Then, following the similar procedure used in Sec.~\ref{sec:SSEA},
one can define a stochastic state vector as $|\psi(t,z^{*},w^{*})\rangle\equiv\langle z,w|\psi_{{\rm tot}}(t)\rangle$,
where we use coherent states $|z\rangle$ and $|w\rangle$ to expand
both bosonic modes $d_{k}$ and $e_{k}$. 
\begin{equation}
\begin{aligned} & \frac{\partial}{\partial t}|\psi(t,z^{*},w^{*})\rangle=\sum_{\substack{\mu,\nu=a,b\\
\mu\ne\nu
}
}\Bigg\{-iH_{\text{A}}+\sigma_{\mu}^{-}z_{\mu t}^{*}+\sigma_{\mu}^{+}w_{\mu t}^{*}\\
 & -\sigma_{\mu}^{+}\int_{0}^{t}ds\left[\alpha_{\mu\mu}(t,s)\frac{\delta}{\delta z_{\mu s}^{*}}+\alpha_{\mu\nu}(t,s)\frac{\delta}{\delta z_{\nu s}^{*}}\right]\\
 & -\sigma_{\mu}^{-}\int_{0}^{t}ds\left[\alpha_{\mu\mu}^{'}(t,s)\frac{\delta}{\delta w_{\mu s}^{*}}+\alpha_{\mu\nu}^{'}(t,s)\frac{\delta}{\delta w_{\nu s}^{*}}\right]\Bigg\}|\psi(t,z^{*},w^{*})\rangle,
\end{aligned}
\label{eq:NM}
\end{equation}
where $z_{\mu t}^{*}=-i\sum_{k}\sqrt{\bar{n}_{k}+1}g_{\mu k}z_{k}^{*}e^{i\omega_{k}t}$
and $w_{\mu t}^{*}=-i\sum_{k}\sqrt{\bar{n}_{k}}g_{\mu k}^{*}w_{k}^{*}e^{-i\omega_{k}t}$
are statistically independent Gaussian noises. $\alpha_{\mu\mu}(t,s)=\sum_{k}{|f_{\mu k}|^{2}e^{-i\omega_{k}(t-s)}}$
and $\alpha_{\mu\mu}^{'}(t,s)=\sum_{k}{|h_{\mu k}|^{2}e^{-i\omega_{k}(t-s)}}$
are the auto-correlation functions, $\alpha_{\mu\nu}(t,s)=\sum_{k}{f_{\mu k}^{*}f_{\nu k}e^{-i\omega_{k}(t-s)}}$
and $\alpha_{\mu\nu}(t,s)=\sum_{k}{h_{\mu k}^{*}h_{\nu k}e^{-i\omega_{k}(t-s)}}$
are the cross-correlation functions with the subindex $\mu\neq\nu$,
where $f_{\mu k}=\sqrt{\bar{n}_{k}+1}g_{\mu k}$ and $h_{k}=\sqrt{\bar{n}_{k}}g_{\mu k}$.
Then, we can replace the functional derivatives in Eq. (~\ref{eq:NM})
by four $O$ operators 
\begin{equation}
\begin{aligned}O_{z\mu}(t,s,z^{*},w^{*})|\psi(t,z^{*},w^{*})\rangle= & \frac{\delta}{\delta z_{\mu s}^{*}}|\psi(t,z^{*},w^{*})\rangle,\\
O_{w\mu}(t,s,z^{*},w^{*})|\psi(t,z^{*},w^{*})\rangle= & \frac{\delta}{\delta w_{\mu s}^{*}}|\psi(t,z^{*},w^{*})\rangle,
\end{aligned}
\end{equation}
and the $O$ operators satisfy the following equations,

\begin{equation}
\begin{aligned}\frac{\partial}{\partial t}O_{z\mu}= & \sum_{\substack{\nu=a,b}
}\Bigg\{[-iH_{\text{A}}+\sigma_{\nu}^{-}z_{\nu t}^{*}+\sigma_{\nu}^{+}w_{\nu t}^{*}-\sigma_{\nu}^{+}\bar{O}_{z\nu}\\
 & -\sigma_{\nu}^{+}\bar{O}_{w\nu},O_{z\mu}]-\sigma_{\nu}^{\dagger}\frac{\delta}{\delta z_{\mu s}^{*}}\bar{O}_{z\nu}-\sigma_{\nu}^{+}\frac{\delta}{\delta z_{\mu s}^{*}}\bar{O}_{w\nu}\Bigg\},\\
\frac{\partial}{\partial t}O_{w\mu}= & \sum_{\substack{\nu=a,b}
}\Bigg\{[-iH_{\text{A}}+\sigma_{\nu}^{-}z_{\nu t}^{*}+\sigma_{\nu}^{+}w_{\nu t}^{*}-\sigma_{\nu}^{+}\bar{O}_{z\nu}\\
 & -\sigma_{\nu}^{+}\bar{O}_{w\nu},O_{w\mu}]-\sigma_{\nu}^{+}\frac{\delta}{\delta z_{\mu s}^{*}}\bar{O}_{z\nu}-\sigma_{\nu}^{\dagger}\frac{\delta}{\delta z_{\mu s}^{*}}\bar{O}_{w\nu}\Bigg\},
\end{aligned}
\end{equation}
with the initial conditions 
\begin{align*}
O_{z\mu}(t,s=t,z^{*},w^{*}) & =\sigma_{\mu}^{-},\\
O_{w\mu}(t,s=t,z^{*},w^{*}) & =\sigma_{\mu}^{+}.
\end{align*}
These equations will help us to fully determine the exact O operator
for the finite-temperature case. Then, similar to the master equation
(\ref{eq:MEQ}), we can also derive the master equation for the finite-temperature
case, 
\begin{equation}
\begin{aligned}\frac{\partial}{\partial_{t}}\rho= & -i[H_{\text{A}},\rho]+\sum_{\mu=a,b}\left([\sigma_{\mu}^{-},\mathcal{M}\{P_{t}\bar{O}_{z\mu}^{\dagger}\}]+{\rm H.c.}\right)\\
 & +\sum_{\mu=a,b}\left([\sigma_{\mu}^{-},\mathcal{M}\{P_{t}\bar{O}_{w\mu}^{\dagger}\}]+{\rm H.c.}\right),
\end{aligned}
\end{equation}
where $P_{t}\equiv|\psi(t,z^{*},w^{*})\rangle\langle\psi(t,z,w)|$
is the stochastic density operator.

\section{Squeezed initial states in waveguide}\label{sec:AppSqueeze}

The general routine of dealing with the squeezed initial states is
already set up in ~\cite{PhysRevA.108.012206}. Considering the initial
state of the waveguide is prepared in squeezed states 
\begin{equation}
|\psi_{\text{W}}(0)\rangle\equiv\bigotimes_{k}\sum_{n_{k}}\frac{(-\tanh r_{k})^{n_{k}}}{\sqrt{\cosh r_{k}}}\frac{\sqrt{(2n_{k})!}}{2^{n_{k}}n_{k}!}|2n_{k}\rangle.
\end{equation}
Assuming the giant atoms and the waveguide are separate at $t=0$,
i.e., $|\psi_{\text{tot}}(0)\rangle=|\psi_{\text{A}}(0)\rangle\otimes|\psi_{\text{W}}(0)\rangle$,
the initial value of the trajectory generally reads $|\psi_{z}\rangle=\langle z|\psi_{\text{tot}}(0)\rangle=\langle z|\psi_{\text{W}}(0)\rangle\otimes|\psi_{\text{A}}(0)\rangle$.
Therefore, the inner product term $\langle z|\psi_{\text{W}}(0)\rangle$
is 
\begin{equation}
\begin{aligned}\langle z|\psi_{W}(0)\rangle & =\prod_{k}\sum_{n_{k}}\frac{z_{k}^{*2n_{k}}}{\sqrt{(2n_{k})!}}\frac{(-\tanh r_{k})^{n_{k}}}{\sqrt{\cosh r_{k}}}\frac{\sqrt{(2n_{k})!}}{2^{n_{k}}n_{k}!}\\
 & =\prod_{k}\frac{1}{\sqrt{\cosh r_{k}}}e^{\frac{-\tanh r_{k}}{2}{(z_{k}^{*})^{2}}}.
\end{aligned}
\end{equation}

Since the inner product $\langle z|\psi_{\text{W}}(0)\rangle\neq1$,
one issue arises that the initial states of trajectories $|\psi_{z}(0)\rangle$
are noise dependent. Thus, we introduce a normalized trajectory $|\varphi_{z}\rangle$,
defined as 
\begin{equation}
|\varphi_{z}\rangle\equiv\frac{|\psi_{z}\rangle}{\prod_{k}\frac{1}{\sqrt{\cosh r_{k}}}e^{\frac{-\tanh r_{k}}{2}(z_{k}^{*})^{2}}},
\end{equation}
to satisfy the coincidence of the initial condition $|\varphi_{z}(0)\rangle=|\psi_{\text{A}}(0)\rangle$.
Consequently, we obtain 
\begin{equation}
\begin{aligned}\frac{\partial}{\partial{z_{k}^{*}}}|\psi_{z}\rangle= & \prod_{k^{'}}\frac{1}{\sqrt{\cosh r_{k}^{'}}}e^{\frac{-\tanh r_{k}^{'}}{2}(z_{k^{'}}^{*})^{2}}\frac{\partial}{\partial{z_{k}^{*}}}|\varphi_{z}\rangle\\
 & +\prod_{k^{'}}\frac{-\tanh r_{k}z_{k}^{*}}{\sqrt{\cosh r_{k}^{'}}}e^{\frac{-\tanh r_{k}^{'}}{2}(z_{k^{'}}^{*})^{2}}|\varphi_{z}\rangle.
\end{aligned}
\end{equation}

Substituting it into the Schr$\ddot{o}$dinger equation, a stochastic
Schr$\ddot{o}$dinger equation of the normalized trajectory $|\varphi_{z}\rangle$
reads

\begin{equation}
\begin{aligned}\frac{\partial}{\partial t}|\varphi_{z}\rangle= & \sum_{\mu=a,b}[-iH_{\text{A}}+z_{\mu t}^{*}\sigma_{\mu}^{-}+w_{-\mu t}^{*}\sigma_{\mu}^{\dagger})\\
 & -i\sigma_{\mu}^{\dagger}\sum_{k}g_{k}e^{-i\omega_{k}t}\frac{\partial}{\partial z_{k}^{*}}]|\varphi_{z}\rangle,
\end{aligned}
\label{eq:psi_z}
\end{equation}
where two stochastic processes are defined as, both in terms of $z_{k}^{*}$,

\begin{equation}
\begin{aligned}z_{\mu t}^{*} & \equiv-i\sum_{k}g_{\mu k}^{*}z_{k}^{*}e^{i\omega_{k}t},\\
w_{\mu t}^{*} & \equiv i\sum_{k}g_{\mu k}z_{k}^{*}e^{i\omega_{k}t}\tanh r_{k},
\end{aligned}
\label{eq:sqnoise}
\end{equation}
respectively. Equation~(\ref{eq:psi_z}) indicates that the new trajectory
$|\varphi_{z}\rangle$ involves two noises. By taking $r_{k}=0$,
it is easy to verify that $w_{t}^{*}=0$, the equations are reduced
to the case without squeezed initial state. In another word, the case
of vacuum initial state is a specific case of Eq. (\ref{eq:psi_z}).

Applying the chain rule, the term of $\frac{\partial}{\partial z_{k}^{*}}|\varphi_{z}\rangle$
can be explicitly extended as a sum of two integrals for the functional
derivatives with respect to the two stochastic processes, that 
\begin{equation}
\begin{aligned} & -i\sum_{k}g_{\mu k}e^{-i\omega_{k}t}\frac{\partial}{\partial z_{k}^{*}}\\
= & -i\sum_{\substack{\mu=a,b\\
\mu\ne\nu
}
}\sum_{k}g_{\mu k}e^{-i\omega_{k}t}\bigg(\int_{0}^{t}ds\frac{\partial z_{\mu s}^{*}}{\partial z_{k}^{*}}\frac{\delta}{\delta z_{\mu s}^{*}}+\int_{-t}^{0}ds\frac{\partial w_{\mu s}^{*}}{\partial z_{k}^{*}}\frac{\delta}{\delta w_{\mu s}^{*}}\\
 & +\int_{0}^{t}ds\frac{\partial z_{\nu s}^{*}}{\partial z_{k}^{*}}\frac{\delta}{\delta z_{\nu s}^{*}}+\int_{-t}^{0}ds\frac{\partial w_{\nu s}^{*}}{\partial z_{k}^{*}}\frac{\delta}{\delta w_{\nu s}^{*}}\bigg)\\
= & -\int_{0}^{t}ds\sum_{k}|g_{\mu k}|^{2}e^{-i\omega_{k}(t-s)}\frac{\delta}{\delta z_{\mu s}^{*}}\\
 & +\int_{-t}^{0}ds\sum_{k}g_{\mu k}^{2}\tanh r_{k}e^{-i\omega_{k}(t-s)}\frac{\delta}{\delta w_{\mu s}^{*}}\\
 & -\int_{0}^{t}ds\sum_{k}g_{\mu k}g_{\nu k}^{*}e^{-i\omega_{k}(t-s)}\frac{\delta}{\delta z_{\nu s}^{*}}\\
 & +\int_{-t}^{0}ds\sum_{k}g_{\mu k}g_{\nu k}\tanh r_{k}e^{-i\omega_{k}(t-s)}\frac{\delta}{\delta w_{\nu s}^{*}}\\
= & -\int_{0}^{t}ds\alpha_{\mu\mu}(t,s)\frac{\delta}{\delta z_{\mu s}^{*}}-\int_{-t}^{0}ds\beta_{\mu\mu}(t,s)\frac{\delta}{\delta w_{\mu s}^{*}},\\
 & -\int_{0}^{t}ds\alpha_{\mu\nu}(t,s)\frac{\delta}{\delta z_{\nu s}^{*}}-\int_{-t}^{0}ds\beta_{\mu\nu}(t,s)\frac{\delta}{\delta w_{\nu s}^{*}},
\end{aligned}
\label{eq:sqCF}
\end{equation}
where $\alpha_{\mu\mu}(t,s)=\sum_{k}|g_{\mu k}|^{2}e^{-i\omega_{k}(t-s)}$
is the correlation function of the noise $z_{\mu t}^{*}$, $\beta_{\mu\mu}(t,s)=\mathcal{M}(z_{\mu t}w_{\mu s}^{*})=\sum_{k}g_{\mu k}{}^{2}e^{-i\omega_{k}(t-s)}\tanh r_{k}$
is the cross-correlation function between the two noises, $\alpha_{\mu\nu}(t,s)=\mathcal{M}(z_{\nu t}w_{\mu s}^{*})=\sum_{k}g_{\mu k}g_{\nu k}^{*}e^{-i\omega_{k}(t-s)}$
and $\beta_{\mu\nu}(t,s)=\mathcal{M}(z_{\mu t}w_{\nu s}^{*})=\sum_{k}g_{\mu k}g_{\nu k}e^{-i\omega_{k}(t-s)}\tanh r_{k}$
are the cross-correlation functions between the two giant atoms. Moreover,
we define two to-be-determined operators $O_{z\mu}(t,s)$ and $O_{w\mu}(t,s)$
as 
\begin{equation}
O_{z\mu}(t,s)|\varphi_{z}\rangle\equiv\frac{\delta}{\delta z_{\mu s}^{*}}|\varphi_{z}\rangle,\quad O_{w\mu}(t,s)|\varphi_{z}\rangle\equiv\frac{\delta}{\delta w_{\mu s}^{*}}|\varphi_{z}\rangle.
\end{equation}
With the $O$ operators, the linear SSE Eq. (\ref{eq:psi_z}) can
be formally written as

\begin{equation}
\begin{aligned}\frac{\partial}{\partial t}|\varphi_{z}\rangle= & (-iH_{\text{A}}+z_{\mu t}^{*}\sigma_{\mu}^{-}+w_{-\mu t}^{*}\sigma_{\mu}^{\dagger})|\varphi_{z}\rangle\\
 & -[\sigma_{\mu}^{\dagger}\bar{O}_{z\mu}^{\alpha}(t)+\sigma_{\mu}^{\dagger}\bar{O}_{w\mu}^{\beta}(t)]|\varphi_{z}\rangle,
\end{aligned}
\end{equation}
where

\begin{align*}
\bar{O}_{z\mu}^{\alpha}(t)\equiv & \int_{0}^{t}ds[\alpha_{\mu\mu}(t,s)O_{z\mu}(t,s)\\
 & +\alpha_{\mu\nu}(t,s)O_{z\nu}(t,s)],\\
\bar{O}_{w\mu}^{\beta}(t)\equiv & \int_{-t}^{0}ds[\beta(t,s)O_{w\mu}(t,s)\\
 & +\beta_{\mu\nu}(t,s)O_{z\nu}(t,s)].
\end{align*}

Here, using the consistency condition that $\frac{\partial}{\partial t}\frac{\delta}{\delta z_{\mu s}^{*}(w_{\mu s}^{*})}|\varphi_{z}\rangle=\frac{\delta}{\delta z_{\mu s}^{*}(w_{\mu s}^{*})}\frac{\partial}{\partial t}|\varphi_{z}\rangle$,
the two operators $O_{z\mu}(t,s)$ and $O_{w\mu}(t,s)$ can be determined
by two evolution equations, respectively: 
\begin{equation}
\begin{aligned}\frac{\partial}{\partial t}O_{z\mu}(t,s)= & \sum_{\substack{\mu,\nu=a,b\\
\mu\ne\nu
}
}\Bigg\{\left[-iH_{\text{A}}+z_{\mu t}^{*}\sigma_{\mu}^{-}+w_{-\mu t}^{*}\sigma_{\mu}^{\dagger},O_{z\mu}(t,s)\right]\\
 & +\left[-\sigma_{\mu}^{\dagger}\bar{O}_{z\mu}^{\alpha}(t)-\sigma_{\mu}^{\dagger}\bar{O}_{w\mu}^{\beta}(t),O_{z\nu}(t,s)\right]\\
 & -\frac{\delta}{\delta z_{\mu s}^{*}}\left(\sigma_{\mu}^{\dagger}\bar{O}_{z\mu}^{\alpha}(t)+\sigma_{\mu}^{\dagger}\bar{O}_{w\mu}^{\beta}(t)\right)\Bigg\},
\end{aligned}
\end{equation}
\begin{equation}
\begin{aligned}\frac{\partial}{\partial t}O_{w\mu}(t,s)= & \sum_{\substack{\mu,\nu=a,b\\
\mu\ne\nu
}
}\Bigg\{\left[-iH_{\text{A}}+z_{\mu t}^{*}\sigma_{\mu}^{-}+w_{-\mu t}^{*}\sigma_{\mu}^{\dagger},O_{w\mu}(t,s)\right]\\
 & +\left[-\sigma_{\mu}^{\dagger}\bar{O}_{z\mu}^{\alpha}(t)-\sigma_{\mu}^{\dagger}\bar{O}_{w\mu}^{\beta}(t),O_{w\nu}(t,s)\right]\\
 & -\frac{\delta}{\delta w_{\mu s}^{*}}\left(\sigma_{\mu}^{\dagger}\bar{O}_{z\mu}^{\alpha}(t)+\sigma_{\mu}^{\dagger}\bar{O}_{w\mu}^{\beta}(t)\right)\Bigg\},
\end{aligned}
\end{equation}
with the initial condition $O_{z\mu}(t,s=t)=\sigma_{\mu}^{-}$ and
$O_{w\mu}(t,s=-t)=\sigma_{\mu}^{\dagger}$.

Moreover, it is worth noting that the initial condition $O_{w\mu}(t,s=-t)=\sigma_{\mu}^{\dagger}$
in the above discussion indicates the nontrivial dynamics of the system.

\subsection*{Spatial-dependent squeezing field}

One of the interesting topic related to squeezed
states in waveguide is the spatial dependence of the squeezing parameter.
Inspired by Ref.~\cite{Gutierrez2023PRR}, if we consider two squeezed
fields (left- and right-propagating) are injected from both ends of
the waveguide, the interference of these two fields creates a standing-wave
pattern of the field quadratures, meaning an atom's position directly
determines whether it probes a maximally squeezed or anti-squeezed
quadrature.

Similar to Ref.~\cite{Gutierrez2023PRR}, the dual
input field can be described by the operators $b_{s}(k)$, where $s=\pm$
represents the propagating directions. Then, the Hamiltonian should
be modified as
\begin{equation}
\begin{gathered}H_{{\rm W}}=\sum_{s=\pm}\int dk\,\omega_{k}b_{s}^{\dagger}(k)b_{s}(k),\\
H_{{\rm int}}=\sum_{\mu=a,b}\sum_{s=\pm}\int dx\,g_{\mu}(x)\left[\sigma_{\mu}^{-}\mathcal{E}_{s}^{\dagger}(x)+{\rm H.c.}\right],
\end{gathered}
\label{eq:ModH}
\end{equation}
where $\mathcal{E}_{\pm}(x)=\int\frac{dk}{\sqrt{2\pi}}b_{s}(k)e^{\pm ikx}$,
and the dispersion relation is assumed linear $\omega_{k}=c|k|$.
The waveguide is initially prepared in a squeezed vacuum state resulting
from injecting broadband squeezed light from both ends. We assume
the squeezed state is characterized by the correlation functions 
\begin{align}
\langle b_{s}^{\dagger}(k)b_{s'}(k')\rangle & =N_{\text{ph}}\,\delta_{s,s'}\delta(k-k'),\\
\langle b_{s}(k)b_{s'}(k')\rangle & =M_{\text{ph}}\,\delta_{s,-s'}e^{i\phi}\delta(k+k'-2k_{c}),
\end{align}
where $k_{c}=\omega_{0}/c$ is the carrier wave vector, $N_{\text{ph}}$
and $M_{\text{ph}}$ satisfy $|M_{\text{ph}}|^{2}=N_{\text{ph}}(N_{\text{ph}}+1)$
for a minimal uncertainty state, and $\phi$ is a reference phase.
The factor $\delta_{s,-s'}$ indicates that squeezing correlates photons
propagating in opposite directions, leading to spatial modulation
of the field quadratures as $\cos[k_{c}(x+x')+\phi]$ \cite{Gutierrez2023PRR}.

Next, we can apply the previously discussed SSE method
for handling squeezed fields here. In the derivation, one may notice
$\langle z|b_{s,k}=z_{s,k}\langle z|$ and $\langle z|b_{s,k}^{\dagger}=\frac{\partial}{\partial z_{s,k}^{*}}\langle z|$,
where the creation and annihilation operators are direction-dependent.

\end{document}